\documentclass[fleqn,usenatbib]{mnras}

\usepackage{newtxtext,newtxmath}

\usepackage[T1]{fontenc}
\usepackage{ae,aecompl}

\usepackage{graphicx}	
\usepackage{amsmath}	
\usepackage{amssymb}	
\usepackage[referable]{threeparttablex}
\usepackage{footnote}
\usepackage[bottom]{footmisc}
\usepackage[normalem]{ulem}

\title[Radio galaxies in LSSs at $z\sim1$]{The Properties of Radio Galaxies and the Effect of Environment in Large Scale Structures at $z\sim1$}

\author[L Shen]{Lu Shen$^{1},$\thanks{Contact e-mail: \href{lushen@ucdavis.edu}{lushen@ucdavis.edu}}
Neal A. Miller$^{2}$,
Brian C. Lemaux$^{1}$,
Adam R. Tomczak$^{1}$,
\newauthor{Lori M. Lubin$^{1}$,
Nicholas Rumbaugh$^{3}$,
Christopher D. Fassnacht$^{1}$,}
\newauthor{Robert H. Becker$^{1}$,
Roy R. Gal$^{4}$, 
Po-Feng. Wu$^{5}$, 
Gordon Squires$^{6}$,}
\\
$^{1}$Physics Department, University of California, Davis, One Shields Avenue, Davis, CA 95616, USA \\
$^{2}$Stevenson University, Department of Mathematics and Physics, 1525 Greenspring Valley Road, Stevenson, MD, 21153, USA\\
$^{3}$National centre for Supercomputing Applications, University of Illinois, 1205 West Clark St., Urbana, IL 61801, USA\\
$^{4}$University of Hawai'i, Institute for Astronomy, 2680 Woodlawn Drive, Honolulu, HI 96822, USA\\
$^{5}$Max-Planck Institut f\"{u}r Astronomie, K\"{o}nigstuhl 17, D-69117, Heidelberg, Germany\\
$^{6}$Spitzer Science centre, California Institute of Technology, M/S 220-6, 1200 E. California Blvd., Pasadena, CA, 91125, USA
}

\date{Accepted 2017 July 31. Received 2017 July 31; in original form 2017 February 14}

\pubyear{2017}

\begin{document}
\label{firstpage}
\pagerange{\pageref{firstpage}--\pageref{lastpage}}
\maketitle

\begin{abstract}
In this study we investigate 89 radio galaxies that are spectroscopically-confirmed to be members of five large scale structures in the redshift range of $0.65 \le z \le 0.96$. Based on a two-stage classification scheme, the radio galaxies are classified into three sub-classes: active galactic nucleus (AGN), hybrid, and star-forming galaxy (SFG). We study the properties of the three radio sub-classes and their global and local environmental preferences. We find AGN hosts are the most massive population and exhibit quiescence in their star-formation activity. The SFG population has a comparable stellar mass to those hosting a radio AGN but are unequivocally powered by star formation. Hybrids, though selected as an intermediate population in our classification scheme, were found in almost all analyses to be a unique type of radio galaxies rather than a mixture of AGN and SFGs. They are dominated by a high-excitation radio galaxy (HERG) population. We discuss environmental effects and scenarios for each sub-class. AGN tend to be preferentially located in locally dense environments and in the cores of clusters/groups, with these preferences persisting when comparing to galaxies of similar colour and stellar mass, suggesting that their activity may be ignited in the cluster/group virialized core regions. Conversely, SFGs exhibit a strong preference for intermediate-density global environments, suggesting that dusty starbursting activity in LSSs is largely driven by galaxy-galaxy interactions and merging.
\end{abstract}




\section{Introduction}
Two main galaxy populations are detected at radio wavelengths: star-forming galaxies (SFGs) and galaxies with an active galactic nucleus (AGN) (e.g.~\citealp{Miley1980, Condon1992, Mauch2007, Smolcic2008, Padovani2009, Padovani2011}). In both cases, the dominant source of the radio emission is synchrotron radiation from relativistic electrons accelerated by supernova or powered by AGN, with a subdominant component of free-free radiation from H II regions (e.g.~\citealp{Condon1992}). 

Radio AGN are typically found in massive quiescent galaxies with older stellar populations~\citep{Miller2002, Best2004, Best2005, Best2007, Mauch2007, Kauffmann2008} and generally retain these properties out to $z \sim 2$~\citep{Malavasi2015} at least for radio galaxy at $log(M_*/M_\odot) > 10$. 
Since radio waves are not affected by dust extinction, as evidenced by the tight far-infrared radio correlation (FRC) observed for dusty SFGs~\citep{Condon1992, Condon2002, Bell2003, Sargent2010, Appleton2014, Magnelli2015}, radio luminosity is widely used as a star formation rate (SFR) indicator (e.g.,~\citealp{Yun2001, Hopkins2003, Bell2003, Bell2005}). The scatter and differences of certain populations in the FRC also allows for the separation of SFGs from AGN (e.g.~\citealp{Padovani2011, Bonzini2013}). 

Radio AGN are often segregated into two sub-populations: high-excitation radio galaxy (HERG) and low-excitation radio galaxy (LERG). These sub-populations are primarily classified based on the presence or absence of high-excitation emission lines in the spectra of their host galaxies~\citep{Hine1979, Laing1994}. Various methods have been applied to separate radio AGN sub-classes by means of radio morphology, the radio luminosity, and optical excitation diagrams (e.g.,~\citealp{Evans2006, Smolcic2009b, Bonzini2013, Padovani2015}). It is found that these two radio AGN populations are different internally; HERGs are powered by radiative-mode (or quasar-mode, cold-mode) accretion, while LERGs are powered by jet-mode (or radio-mode, hot-mode) accretion (e.g.~\citealp{Ciotti2010, Best2012}). In the former case, the accretion is efficient near the Eddington rate, while in the latter case the accretion is considerably less efficient. Additionally, the black holes that they contain typically have very different masses, with the former population hosting black holes that are, on average, approximately an order of magnitude less massive~\citep{Hickox2009, Best2012}.

Environments, another subject analyzed in the paper, play an important role on triggering and quenching processes. For SFGs, this role is evidenced by the relationships of galaxy colour, morphology, stellar mass, and star formation rate with various measures of environment (e.g.,~\citealp{Dressler1980, Peng2010, Grutzbauch2011, Peng2012, Wetzel2013}). However, it is not fully understood which process accounts for these relationships. It has been proposed that ram pressure stripping, harassment, and tidal forces would act with different efficiencies and over different time scales, depending on the properties of the galaxies and on their environments (e.g.,~\citealp{Byrd1990, Abadi1999, Bahe2013, Merluzzi2016}). 
Meanwhile, AGN may be largely triggered through galaxy merging or strong tidal interactions (e.g.~\citealp{Kauffmann2000, Hopkins2006, Hopkins2008}), though other environmental or secular processes may also be responsible (\citealp{Kocevski2012}, and reference therein).
Resultant AGN activity can subsequently either trigger (positive feedback) or suppress (negative feedback) star formation activity (e.g.,~\citealp{Mullaney2012, Zubovas2013, Lemaux2014, Hickox2014, King2015})

Radio AGN are found preferentially near or in clusters/groups at low redshift (e.g.,~\citealp{Miller2002, Best2004, Argudo2016}) with such a preference appearing to persist up to $z \sim 2$ (e.g.,~\citealp{Magliocchetti2004, Lindsay2014, Hatch2014, Malavasi2015}). 
As such, radio AGN are widely used to detect protoclusters at $z > 1$ (e.g.~\citealp{Best2000, Venemans2007, Wylezalek2013}). HERGs and LERGs also show different environmental preferences, with the former restricted to low density environments and the latter occupying a wider range of densities (e.g.,~\citealp{Donoso2010, Gendre2013}). Radio SFGs, conversely, tend to be located in lower global and local density environments at low redshift (e.g.,~\citealp{Miller2002, Best2004}).

In this study, we investigate 89 radio galaxies that are spectroscopically confirmed members of five large-scale structures (LSSs) at $z\sim1$ drawn from the Observations of Redshift Evolution in Large Scale Environments (ORELSE; ~\citealp{Lubin2009}) survey. 
The ORELSE survey is a systematic search for large-scale structures (LSSs) around an original sample of 20 galaxy clusters in a redshift range of $0.6 < z < 1.3$ with the field of view of all fields $\sim 0.25-0.5 deg^2$. The goal is to study galaxy properties over a wide range of local and global environments. 
In this paper, we use Very Large Array (VLA) 1.4GHz imaging to locate radio galaxies by matching them to a large data set of spectroscopically-confirmed members and separate them into three sub-classes: AGN, Hybrid and SFG. We compare the properties of their hosts galaxies in colour, stellar mass, and radio luminosity, and study their environmental preferences. We also compare such environmental preferences of the three radio sub-classes with a control sample of spectroscopically-confirmed members. For the purpose of studying environmental effects on radio galaxies, rather than environmental properties driven solely by colour or stellar mass, we carefully select a non-radio control sample matched in the stellar mass and rest-frame colour for each radio sub-class and perform a comparison on the global and local environments.

The paper is outlined as follows. Section~\ref{sec:lss} introduces the five LSSs. In Section~\ref{sec:obs} we discuss the observational data. We propose a classification scheme in Section~\ref{sec:classification} to divide radio galaxies into AGN, Hybrids, and SFGs. In Section~\ref{sec:properties} we analyze the properties of the host galaxies of the three radio galaxy sub-classes and their environmental preferences. We then, in Section~\ref{sec:discussion}, discuss the results of the three radio sub-classes separately and lead to a scenario for each sub-class. Finally, in Section~\ref{sec:conclusion}, we summarize all our results.
Throughout this paper all magnitudes, including those in the IR, are presented in the AB system \citep{Oke1983, Fukugita1996}. All equivalent width measurements are presented in the rest frame and we adopt the convention of negative equivalent widths corresponding to a feature observed in emission. All distances are quoted in proper units. We adopt a concordance $\Lambda$CDM cosmology with $H_0 = 70~km~s^{-1} Mpc^{-1}$, $\Omega_{\Lambda} = 0.73$, and $\Omega_{M} = 0.27$, and a Chabrier initial mass function (IMF;~\citealp{Chabrier2003}).

\section{The ORELSE Large-Scale Structure Sample}
\label{sec:lss}
In this section, we give a picture of the five large-scale structures (LSSs) in our sample. These five fields are the original fields from the ORELSE survey that were observed with coordinated VLA and Chandra observations (see~\citealp{Rumbaugh2012} for details on the latter). They span $7\sim15$ Mpc in the plane of the sky and $50\sim250$ Mpc along the line of sight. The five LSSs are defined based on both the line of sight direction (as determined from the velocity offsets) and the projected distance to the nearest cluster/group. We, here, rely on~\citet{Rumbaugh2012, Rumbaugh2016} for the LSS redshift boundaries, determined by visually examining their redshift histogram, and designed to include all galaxies in each overall LSS. Meanwhile, the spectroscopic covered typically extends to 5 Mpc from any given cluster/group within the redshift range of each LSS.
For these five fields, we have fully reduced radio catalogs (of the 16 fields in ORELSE with VLA/JVLA imaging) accompanying fully reduced photometric and spectroscopic catalogs. Moreover, they span the full range of ORELSE structures, in terms of halo mass and dynamics, albeit without the higher redshift structures which will be included in future works. We summarize the central position, the number of clusters or groups, and the range of velocity dispersions in each field, see Table~\ref{tab:lss}, while, the spectroscopy properties will be described in Section~\ref{sec:specobs}. 

\subsection{SC1604}
The SC1604 supercluster was originally discovered in~\citet{Gunn1986} as a cluster candidate, and then confirmed by~\citet{Oke1998}.
The LSS of SC1604 spans a redshift range of $0.84 < z < 0.96$, consisting of 5 clusters and 3 groups, with a cluster being $\sigma_\nu > 550$ km/s. Many detailed studies have been performed on this structure~\citep{Lubin1998a, Lubin2000, Postman2001, Best2002, Gal2004, Rieke2004, Postman2005, Homeier2006, Gal2008, Kocevski2009a, Lemaux2010, Lemaux2011,  Lemaux2012, Kocevski2011, Rumbaugh2012,  Rumbaugh2013, Ascaso2014, Wu2014, Rumbaugh2016}. 

\subsection{SG0023}
The SG0023 structure was originally discovered in~\citet{Gunn1986}, and confirmed including two additional groups by~\citet{Oke1998}.
The LSS spans the redshift range of $0.82 < z < 0.87$, consisting of five merging galaxy groups~\citep{Lubin2009, Lubin1998a, Lemaux2016}. Many additional studies have been performed on this structure~\citep{Postman1998,  Lubin1998b, Best2002, Rumbaugh2012,  Rumbaugh2013, Lemaux2016, Rumbaugh2016}. 

\subsection{SC1324}
The SC1324 supercluster was originally discovered in~\citet{Gunn1986}, with two clusters confirmed. Subsequent studies have increased the structure to 3 clusters and 4 group, spanning the redshift range of $0.65 < z < 0.79$~\citep{Postman2001, Lubin2002, Lubin2004, Rumbaugh2012, Rumbaugh2013, Rumbaugh2016}. 

\subsection{RXJ1757}
The RXJ1757.3+6631 cluster (hereafter RXJ1757, described in \citealp{Rumbaugh2012}), was discovered as part of the \textit{ROSAT} North Ecliptic Pole (NEP) survey, and identified as NEP200~\citep{Gioia2003}, and consisting of only one cluster at $z = 0.69$. Because we detect likely substructures associated with RXJ1757, we consider this system to be a LSS. This LSS spans the redshift range of $0.68 < z < 0.71$, and has only limited studies~\citep{Rumbaugh2012, Rumbaugh2013}. 

\subsection{RXJ1821}
\label{sec:RXJ1821}
The RXJ1821.6+6827 cluster (hereafter RXJ1821; described in \citealp{Lubin2009} and \citealp{Rumbaugh2012}), was discovered in the \textit{ROAST} NEP survey, and identified as NEP5281, consisting of only one cluster at $z = 0.8168$. Similarly, we likely find surrounding substructure associated with RXJ1821. This LSS, spanning the redshift range of $0.80 < z < 0.84$, has been studies by~\citet{Gioia2004, Lubin2009, Lemaux2010, Rumbaugh2012, Rumbaugh2013, Rumbaugh2016}.

\begin{table*}
	\caption{Properties of ORELSE large-scale structure}
	\label{tab:lss}
	\begin{threeparttable}
	    \begin{tabular}{lccccccc} 
		    \hline
		    \hline
	    	Field & R.A.\tnote{1} & Decl.\tnote{1} & $<z_{spec}>$\tnote{2} & Confirmed  & $\sigma$  & $z_{spec} $ & Comfirmed \\
	    	& (J2000) & (J2000)  &  & Clusters/Groups\tnote{3} & Range\tnote{4}  & LSS Range & Members\tnote{5}\\
		    \hline
		    SC1604 & 16:04:15 & +43:21:37 & 0.90 & 8 & 300-800 & 0.84$\sim$0.96 & 520 \\
	    	    SG0023 & 00:23:52 & +04:22:51 & 0.84  & 5 & 200-500 & 0.82$\sim$0.87 & 246\\
		    SC1324 & 13:24:35 & +30:18:57 & 0.76  & 4 & 200-900 & 0.65$\sim$0.79 & 421 \\
		    RXJ1757 & 17:57:19.4 & +66:31:29 & 0.69 & 1 & 862.3$\pm$107.9 & 0.68$\sim$0.71 & 75 \\
		    RXJ1821 & 18:21:32.4 & +68:27:56 & 0.82  & 1 & 1119.6$\pm$99.6 & 0.80$\sim$0.84 & 129 \\ 
		    \hline
	    \end{tabular}
	    \begin{tablenotes}
		\item[1] Coordinates for SC1604, SG0023, SC1324 are the median of central positions of clusters/groups, while RXJ1757 and RXJ1821 are given as the centroid of the peak of diffuse X-ray emission associated with the respective cluster.  
		\item[2] Average $z_{spec}$ of member galaxies in each LSS. 
        \item[3] Clusters/Groups are spectroscopically confirmed using the method presented in \citet{Gal2008}.
		\item[4]  Galaxy line-of-sight velocity dispersion in units of km s$^{-1}$, measured within 1 Mpc. For SC1604, SG0023, SC1324, the range of velocity dispersions of clusters/groups are given. For RXJ1757 and RXJ1821 the velocity dispersion of the single clusters are given.
		\item[5] Number of spectroscopic objects confirmed in the LSS redshift range, with quality flag Q = 3 or 4, $18.5 \le i' \le 24.5$ (for more details see Section~\ref{sec:specobs}).
	    \end{tablenotes}
    \end{threeparttable}
\end{table*}
\section{Observations and Reductions}
\label{sec:obs}
In this section we present the observations and the reduction of radio, photometric, and spectroscopic data. We also describe the optical matching technique used to confirm radio galaxies with spectroscopic counterparts.
By the end of this section, we confirm 89 radio galaxies in the five LSSs.
\subsection{Radio Observations}
\label{sec:radioobs}	
Each of the five LSSs were observed in 2006-2009 in the same manner using the VLA at 1.4GHz in its B configuration, where the resulting FWHM resolution of the synthesized beam is about 5$^{\prime\prime}$. 
Given the approximate $31^\prime$ diameter field of view (i.e., the FWHM of the primary beam), we opted to observe each target using a single pointing of the VLA with the exception of SC1324 for which two pointings were used to cover the full extent of the structure. 
As the observations were completed prior to the upgrade of the VLA, we used the ``4 mode" correlator setting which consists of seven 3.125~MHz channels at each of two intermediate frequencies. This mode was the best compromise for sensitivity (from a wider total bandwidth) with lesser distortion of the beam (the smaller individual channels reduces bandwidth smearing). 
Each LSS was observed over the course of several days using individual tracks of several hours per day to insure good $(u,v)$-coverage with a nearly circular synthesized beam. Net integration times were chosen to result in final $1\sigma$ sensitivities of about 10$\mu$Jy per beam. Details of the observations may be found in Table~\ref{tab:radio_obs}. 

In general, the procedures for calibration and imaging of the data paralleled those presented in~\citet{Miller2013} except that the LSS data correspond to a single or, at most, two pointings per target field and required substantially fewer facets to handle the effects of sky curvature. 
The observations on each date included a gain calibrator (3C48 for SC0023, 3C286 for all other LSSs) along with a nearby (ranging from $3.4^\circ$ to $7.6^\circ$) phase calibrator. 
The $(u,v)$ data for these calibrators were edited to remove obvious interference and other aberrational data, and the resulting gain and phase calibrations were applied to the target fields. 
The target fields were imaged using a ``flys-eye" pattern of seven facets, with each facet having 1024 1.5$^{\prime\prime}$ pixels per side. Wide-field maps made at lower resolution identified bright sources outside this primary-beam covering flys-eye, and these sources along with bright sources inside the flys-eye received their own dedicated smaller facets for imaging. 
For each specific observation date, $(u,v)$ data sets were generated and self-calibrated. 
The $(u, v)$ data were then averaged to produce a single $(u,v)$ data file corresponding to the complete set of observations (i.e., combining all observation dates) for each target LSS field. 
Final rounds of imaging and self-calibration were then performed on these $(u,v)$ data. In these final imaging steps, the data for a specific LSS target were imaged separately by intermediate frequency, polarization, and in roughly 2-hour blocks in  Local Sidereal Time (LST). The resulting images were then combined using variance weighting to arrive at the final images for each target LSS field. 

The final images were then used to generate source catalogs. 
The NRAO's Astronomical Image Processing System (AIPS) task SAD created the initial catalogs by examining all possible sources having peak flux density greater than three times the local RMS noise. We then instructed it to reject all structures for which the Gaussian fitted result has a peak below four times the local RMS noise. Because Gaussian fitting works best for unresolved and marginally resolved sources, residual images created by SAD having subtracted the Gaussian fits from the input images were inspected in order to adjust the catalog. This step added those extended sources poorly fitted by Gaussians. 
Peak flux density, integrated flux density, and their associated  flux density errors ($\sigma$) are generated by SAD. We use the peak flux density unless the integrated flux is larger by more than $3\sigma$ than the peak flux for each individual source. 

\begin{table*}
	\centering
	\caption{VLA Radio Observation Characteristics}
	\label{tab:radio_obs}
	\begin{threeparttable}
	    \begin{tabular}{lclcccc} 
		    \hline
		    \hline
	    	    Field  & Code  & Observation Dates\tnote{1} & R.A.\tnote{2} & Decl.\tnote{2} & RMS\tnote{3} & Beam \\
	    	   	 &  	&  	& (J2000) & (J2000) & ($\mu$Jy) & Size \\
		    \hline
		    SC1604  & S7218  & 07/07/06, 07/10/06, 07/13/06, 09/04/06 & 16:04:13.2 & +43:13:51  &   9.3 & $5.3"\times 5.0"$\\
		    SG0023 & S8597 & 11/06/07, 11/07/07, 11/09/02, 11/10/07 &   00:23:50.7  & +04:22:45  & 13.9 & $4.4"\times 4.4"$\\
		   SC1324N & S9484 & 02/18/09, 02/19/09, 02/25/09, 02/26/09, 03/02/09, 03/04/09 &13:24:49.5 & +30:51:34 & 11.4  &  $4.7"\times 4.5"$ \\
		   SC1324S  &  S9484 & 02/18/09, 02/19/09, 02/25/09, 02/26/09, 03/02/09, 03/04/09 & 13:24:42.5 & +30:16:30 & 10.5  & $4.7"\times 4.5"$ \\
		   RXJ1716   & SA598  & 02/23/09, 02/27/09, 02/28/09, 03/01/09  &   17:57:19.8 & +66:31:39  & 10.5  & $4.9"\times 4.7"$\\
		   RXJ1821  &  SA598  & 02/23/09, 02/27/09, 02/28/09, 03/01/09  &   18:21:38.1 & +68:27:52  & 9.3  &  $4.9"\times 4.7"$\\
		    \hline
	    \end{tabular}
	    \begin{tablenotes}
	    	\item[1] For SC1604, this is starting day of the observations, while for other fields, these are the date when the observation was taken;
		\item[2] Spatial position for the phase centre of observation.
		\item[3] RMS sensitivity for final image associated with all data for that pointing.
	    \end{tablenotes}
    \end{threeparttable}
\end{table*}

\subsection{Photometric Observations and Catalogs}
\label{sec:photoobs}
Comprehensive photometric catalogs are constructed for all five LSSs. We summarize the available optical and near-infrared (NIR) observations for each LSS and the reduction of these data in Table~\ref{tab:images1} and~\ref{tab:images2} . A full description of the reduction process will be given in Tomczak et al (2017, \textit{in prep.}). 
Optical imaging was taken with the Large Format Camera (LFC;~\citealp{Simcoe2000}) on the Palomar 5-m telescope, using Sloan Digital Sky Survey (SDSS, \citealp{Doi2010})-like r$^{\prime}$, i$^{\prime}$ and z$^{\prime}$ filters, reduced in the Image Reduction and Analysis Facility (IRAF,~\citealp{Tody1993}), following the method in~\citet{Gal2008}. We also use R, I, and Z band optical imaging from Suprime-Cam~\citep{Miyazaki2002} on the Subaru 8-m telescope, reduced with the \texttt{SDFRED2} pipeline~\citep{Ouchi2004} supplemented by several routines Traitement \'El\'ementaire R\'eduction et Analyse des PIXels (\texttt{TERAPIX})\footnote{http://terapix.iap.fr}. 
Some J and K band data was taken with the United Kingdom Infrared Telescope Wide-Field Camera (WFCAM;~\citealp{Hewetti2006}) mounted on the United Kingdom Infrared Telescope (UKIRT) and was reduced using the standard UKIRT processing pipeline courtesy of the Cambridge Astronomy Survey Unit\footnote{http://http://casu.ast.cam.ac.uk/surveys-projects/wfcam/technical}. 
Additional, J and Ks band imaging was taken using the Canada-France-Hawaii Telescope Wide-field InfraRed Camera (WIRCam;~\citealp{Puget2004}) mounted on the Canada-France-Hawai'i Telescope (CFHT) and was reduced through the I'iwi preprocessing routines and \texttt{TERAPIX}. Infrared imaging at 3.6, 4.5, 5.8, and 8.0 $\mu$m (5.8 and 8.0 $\mu$m only available for SC1604) was taken using the Spitzer telescope Infrared Array Camera (IRAC;~\citealp{Fazio2004}). The basic calibrated data (cBCD) images provided by the Spitzer Heritage Archive were reduced using the MOsaicker and Point source EXtractor (MOPEX; \citealp{Makovoz2005}) package augmented by several custom Interactive Data Language (IDL) scripts written by J. Surace.

Photometry was obtained by running Source Extractor (SExtractor;~\citealp{Bertin1996}) on point spread function (PSF)-matched images convolved to the image with the worst seeing. Magnitudes were extracted in fixed-apertures to ensure that the measured colours of galaxies are unbiased by different image quality from image to image. 
Also, the package T-PHOT~\citep{Merlin2015} was used for Spitzer/IRAC images, due to the large point spread function of these data that can blend profiles of nearby sources together and contaminate simple aperture flux measurements. See \citet{Lemaux2016} for more details. 

Spectral Energy Distribution (SED) fitting was performed in three stages. At the first stage, aperture magnitudes were input to the code Easy and Accurate $z_{phot}$ from Yale (EAZY,~\citealp{Brammer2008}). This includes 6 templates derived from a non-negative matrix factorization decomposition of the Project d'\'Etude des GAlaxies par Synth\`ese E\'volutive (PE\'GASE;~\citealp{Fioc1997}) template library with one additional template from \citet{Maraston2005} representing an old stellar population. This template set is generally more effective at identifying the location of strong spectral breaks in broadband photometry than using individual templates. We refer the reader to~\citet{Brammer2008} for a more thorough discussion of the templates used to fit for photometric redshifts. 

For each object, the $z_{phot}$ is estimated from a probability distribution function (PDF) (hereafter P(z)), generated by calculating the $\chi^2$ of the photometries with respect to PE\'GASE models.  
A separate set of fitting was performed utilizing a library of stellar templates~\citep{Pickles1998} to determine stars from galaxies. 
In the second stage of the fitting process, the code EAZY was run again using $z_{phot}$ or high quality spectroscopic redshifts when available to derive rest-frame magnitudes for all photometric objects. Magnitudes were corrected to infinity by applying the difference between MAG\_APER and MAG\_AUTO in the detection band to all magnitudes when driving physical parameters from SED fitting. 
As for the final stage, SED fitting on aperture corrected magnitudes was performed with the Fitting and Assessment of Synthetic Templates (FAST;~\citealp{Kriek2009}) code, with the same redshift used in the second stage. 
Because the templates used with EAZY cannot be used to estimate physical properties such as stellar mass we employ the SPS library of~\citet{Bruzual2003} assuming a~\citet{Chabrier2003} stellar initial mass function (IMF) and solar metallicity for this part of the analysis. For dust extinction we adopt the ~\citet{Calzetti2000} attenuation curve. We adopt delayed exponentially declining star-formation histories ($\mathrm{SFR} \propto t \times e^{-t / \tau}$), allowing log($\tau/\mathrm{yr}$) to range between 8.5-10 in steps of 0.5, log($age/\mathrm{yr}$) to range between 8-10 in steps of 0.2, and $\mathrm{A_V}$ to range between 0-3 in steps of 0.1. 
Only the stellar mass and V-band attenuation in magnitudes (Av) derived from this fitting are used in this paper. It is important to note that the parameter space used in our stellar population synthesis (SPS) modeling is smaller than in other contemporary studies. Nevertheless, we have tested our analysis using stellar masses derived from a more extended parameter space and find that our results remain unchanged.

The precision and accuracy of the photometric redshifts were estimated from fitting a Gaussian to the distribution of $(z_{spec} - z_{phot})/(1 + z_{spec})$ measurements in the range $0.5 \le z \le 1.2$ and was found to be $\sigma_{\Delta z/(1+z)} \sim 0.03$ with a catastrophic outlier rate ($\Delta z/(1 + z) > 0.15$) of $5-10\%$ for all five fields to a limit of $i' \le 24.5$. A slight systematic offset from zero ($\Delta z/(1 + z_{phot}) \sim 0.01$) was noticed for all five fields. The value of this offset, multiplied by ($1+z_{phot}$), was applied to all raw $z_{phot}$ values (See \citet{Lemaux2016}, and Tomczak et al. (\textit{in prep}) for more details). 
For galaxies with known or suspected AGN in our sample based on our radio classification, we check their stellar template fitting. They all provide good fits to the observed photometry, thus $M_*$ and Av would not be different if AGN/Hybrid templates were instead used. 

We make use of {\it Spitzer} MIPS 24$\mu$m observations for estimating rest-frame total infrared luminosities (8-1000$\mu m$, $L_{\mathrm{TIR}}$), which are available for SC1604, SG0023, and RXJ1821.
Due to the large point spread function of these data (PSF $\gtrsim$ 4"), profiles of nearby sources would blend together and contaminate simple aperture flux measurements.
To help mitigate this effect we employ T-PHOT~\citep{Merlin2015}. In brief, the methodology of T-PHOT is as follows. First, the positions and morphologies of objects are obtained for use as priors based on a segmentation map produced by Source Extractor~\citep{Bertin1996}) of a higher-resolution image.
Cutouts are taken from this higher-resolution image which are then used to create low-resolution 24$\mu$m models of each object by smoothing with a provided convolution kernel.
These models are then simultaneously fit to the 24$\mu$m image until optimal scale factors for each object are obtained as assessed through a global $\chi^2$ minimization.
We run T-PHOT in ``{\it cells-on-objects}" mode where when fitting a model for a given object only neighbours within a $1^{\prime}\times1^{\prime}$ box centreed on the object are considered in the fitting.
After this initial sequence, T-PHOT is then rerun in a second pass in which registered kernels generated during the first pass are utilized to account for mild astrometric differences between the input images.
Output fluxes are the total model flux from the best fit.

To estimate $L_{TIR}$, we make use of the infrared spectral template introduced by~\citet{Wuyts2008}. 
This template is constructed by averaging the logarithm of all templates in the \citet{Dale2002} infrared spectral library.
For each galaxy we scale the infrared spectral template, shifted to the galaxy's redshift ($z_{\mathrm{spec}}$ when available) to obtain its total 24$\mu$m flux estimate.
We then shift the template back to the rest-frame and integrate between 8-1000$\mu$m treating this as the bolometric infrared luminosity, $L_{\mathrm{TIR}}$.
A luminosity-independent conversion of flux to $L_{\mathrm{TIR}}$ was first suggested at by the stacking results of \citet{Papovich2007}.
Since then this technique has been supported by several studies across a broad range of redshifts and stellar masses (see e.g. \citealp{Muzzin2010, Wuyts2011, Tomczak2016}).

We apply a quality flag for each photometric detection to include detections above $3\sigma$ in the detection image with coverage in at least 5 images, and exclude stars, saturations\footnote{Defined as cases where more than 20\% of the pixels in the photometric aperture are saturated in the detection image} and bad SED fitting ($\sim$ 1$\%$ worst $\chi^2$ values). Photometric objects are also restricted to be in the same spatial area as the spectroscopic observation in each field and observed $i'$ band $18.5 \le i' \le 24.5$ to maximize the completeness of spectroscopy, while still including the vast majority of our spectroscopically-confirmed members. 

Moreover, we use a wider LSS photometric redshift range for photometric objects without confirmed $z_{spec}$, due to the uncertainty of $z_{phot}$, which varies from field to field. We use $z_{phot}$ selection described in Appendix~\ref{app:spec-photo} to select the photometric members (see Table~\ref{tab:photo} for number of photometry in the LSSs). The $z_{phot}$ redshift range defined using Equation~\ref{eq:photoz}. 

\subsection{Spectroscopic Observations}
\label{sec:specobs}

\begin{table*}
	\caption{DEIMOS Spectroscopic Observation Characteristics}
	\label{tab:obs}
	\begin{threeparttable}
	    \begin{tabular}{lllllcc} 
		    \hline
		    \hline
	    	LSS & Central & Approx. Spectral & Exp. Time & Avg. Seeing & Num. of  & Num. of  \\
	    	& $\lambda$ ({\AA}) & Coverage ({\AA})  &  Range (s) & Range (") & Mask & Targets\tnote{1} \\
		    \hline
		    SC1604 & 7700 & 6385-9015 & 3600-14400 & 0.50-1.30 & 18 & 2445\\
	    	    SG0023 & 7500-7850 & 6200-9150 & 5700-9407 & 0.45-0.81 & 9 & 902\\
		    SC1324 & 7200 & 5900-8500 & 2700-10800 & 0.44-1.00 & 12 & 1237\\
		    RXJ1757 & 7000-7100 & 5700-8400 & 6300-14730 & 0.47-0.82 & 6 & 728\\
		    RXJ1821 & 7500-7800 & 6200-9100 & 7200-9000 & 0.58-0.86 & 6 &593\\ 
		    \hline
	    \end{tabular}
	    \begin{tablenotes}
		\item[1] Including spectroscopic target with $18.5 \le i' \le 24.5$, including sirendips. 
	    \end{tablenotes}
    \end{threeparttable}
\end{table*}

Spectroscopic targets were selected based on the optical imaging in the $r'$, $i'$, and $z'$ from LFC, following the methods in~\citet{Lubin2009}. 
We targeted objects with priority given to redder galaxies at the presumed redshifts of the LSSs, which was intended to perferentially confirm red sequence members over bluer counterparts, generally including galaxies with an $i'$-band magnitude brighter than 24.5. 
In addition, for certain masks we prioritized X-ray and radio detected objects.  
The optical spectroscopy was primarily taken with the Deep Imaging and Multi-Object Spectrograph (DEIMOS;~\citealp{Faber2003}) on the Keck II 10m telescope. 
DEIMOS is an efficient instrument for a spectroscopic survey of high-redshift large-scale structures, because of its large field of view ($16.'7 \times 5.'0$) and its capability of positioning up to 120+ galaxies per slitmask. 
We used the 1200 line mm$^{-1}$ grading with $1"$-wide slit at various central wavelengths (see Table~\ref{tab:obs}), with wavelength coverage of approximately 2600{\AA}. 
Total exposure times are 1-4.5 hours per mask, which varied based on observational conditions and the magnitude distributions of targets. A few additional redshifts from the Low Resolution Imaging Spectrometer (LRIS; \citealp{Oke1995}) were added for SC1604, SG0023 and RXJ1821 (see~\citealp{Oke1998, Gal2004, Gioia2004}).   

During the reduction of DEIMOS data, serendipitous detections (hereafter serendips) were identified by eye in the spatial profile of slits. If such detections existed, a serendipitous spectrum was generated using the method described in~\citet{Lemaux2009}. Redshifts and quality flag (Q) were determined for all targets serendipitous detections. See~\citet{Gal2008, Newman2013} for more details on the quality flags. We use spectroscopic objects only with Q = 3, 4 in this paper. Q = 3 means one secure and one marginal feature were used to derive the redshift.  Targets of Q = 4 are at least two secure features were used.  Q = -1 flag means the target was definitely determined to be star. Q = 1 or Q = 2 indicated that we could not determine a secure redshift. Moreover, the same $i'$ band magnitude use flag  ($18.5 \le i' \le 24.5$), as  applied to the photometric catalogs, was applied to the spectroscopic samples , to keep these two catalogs consistent. We list the number of spectroscopically-confirmed members, within the LSS redshift range in Table~\ref{tab:lss}. 

\subsection{Optical Matching}
\label{sec:matching}

\begin{table*}
	\centering
	\caption{Number of Radio sources, and Photometric and Spectroscopic Matches}
	\label{tab:matches}
	\begin{threeparttable}
	    \begin{tabular}{lcccccc} 
		    \hline
		    \hline
	    	    Field & Radio & Radio-phot & Radio-spec & Radio-phot & Radio & Radio\\
	    	    & Sources\tnote{1} & Matches\tnote{2} & Matches\tnote{3} & Sample\tnote{4} & Confirmed\tnote{5} & Extra\tnote{6}\\
		    \hline
		    SC1604 & 501 & 193 & 109 & 42 & 32 & 9\\
	    	    SG0023 & 313 & 95 & 53 & 12 & 6 & 2\\
		    SC1324 & 565 & 165 & 36 & 49 & 21 & 4+1\\
		    RXJ1757 & 258 & 91 & 45 & 6 & 3 & 0\\
		    RXJ1821 & 201 & 45 & 29 & 13 & 9 & 3 \\ 
		    \hline
	    \end{tabular}
	    \begin{tablenotes}
	    \item[1] Radio sources with a significance $> 4\sigma$ in at least one of the radio integrated or peak flux;
		\item[2] Radio sources matched to photometric counterparts in the field, restricted with $18.5 \le i' \le 24.5$ and the same spatial area as corresponding spectroscopic imaging; 
		\item[3] Radio sources matched to spectroscopic counterparts in the field, restricted with $18.5 \le i' \le 24.5$ and qualify flag;
		\item[4] Radio objects that have counterparts of photometric members in the LSSs in $\sigma_j =1\arcsec$. For photometric members see explanation in Section~\ref{sec:photoobs}; 
		\item[5] Radio objects matched in $\sigma_j =1\arcsec$ that have spectroscopically-confirmed counterparts in the LSSs. 
        \item[6] Radio extra confirmed members in $\sigma_j =2\arcsec + $confirmed radio double. The final number of radio confirmed sample is adding the last two columns. 
	    \end{tablenotes}
    \end{threeparttable}
\end{table*}

Prior to beginning the full optical-radio matching process, the relative astrometry of the two sets of imaging was compared in the following way. The positions of radio point sources with a detection significance larger than 10 in each field were drawn from the catalogs generated in Section~\ref{sec:radioobs} and were nearest-neighbour matched to the photometric catalog in each corresponding field. The median offsets of the radio sources relative to their nearest-neighbour optical counterparts in $\alpha$ and $\delta$ were measured for each individual field. The offsets were always less than 0.4$\arcsec$ and typically much less. The inverse of these offsets were subsequently applied to the full radio catalog for each field. The resultant normalized median absolute deviation \citep{Hoaglin1983} of the corrected radio positions of the $>10\sigma$ sources relative to those of their optical counterparts was found to be 0.25$\arcsec$, which is considerably smaller than the radius ($1\arcsec$) chosen for optical-radio matching.

To search for radio galaxies within the individual large-scale structures, we match our radio sources to the photometric catalogs. We use the maximum likelihood ratio technique described in \citet{Rumbaugh2012}, which was developed by~\citet{Sutherland1992} and also used by~\citet{Taylor2005, Gilmour2007, Kocevski2009a}. The main statistic calculated in each case is the likelihood ratio (LR), which estimates the probability that a given optical source is the genuine match to a given radio source relative to the arrangement of the two sources arising by chance. The LR is given by the equation
\begin{equation}
    LR_{i,j} = \frac{w_i exp(-r^2_{i,j}/2\sigma^2_j)}{\sigma^2_j}
	\label{eq:LR}
\end{equation}
Here, $r_{i,j}$ is the separation between objects i and j, $\sigma_j$ is the positional error of object j, where we use $1\arcsec$ as positional error of all radio detection~\citep{Condon1997}, and $w^2_i = 1/n(< m_i)$ is the inverse of the number density of optical sources with magnitude fainter than the observed $i'$ band magnitude. The inclusion of the latter quantity is designed to weight against matching to fainter optical objects that are more likely to have chance projections. For each radio source, we carried out a Monte Carlo (MC) simulation to estimate the probability that each optical counterpart is the true match using the LRs. We adopt the same threshold for matching to a single or double objects as~\citet{Rumbaugh2012}. 

The optical matching is done to the overall photometric catalogs, aimed to increase the probability of the true match. We obtain a total 589 radio-phot matches in the maximum likelihood matching. In this paper we mainly analyze radio objects which have photometric counterparts with spectroscopically-confirmed redshifts (refereed to as radio-spec matches) and within the LSS redshift range (see Table~\ref{tab:lss}), referred to as radio confirmed galaxies. We obtain 272 radio-spec matches from the radio-phot matches, 71 of which are members of the LSSs (see Table~\ref{tab:matches} for numbers of radio-phot matches, radio-spec matches and radio confirmed galaxies in each LSS).

Radio sources may have extended morphology, however, in the maximum likelihood ratio technique, each radio source has been assumed to be a point source coincident with the optical center. 
To account for astrophysical and astrometric offsets, we use $\sigma_j = 2\arcsec$ as the search radius and ran the optical matching again. We examined the additional matches by overlaying radio flux contours for each match on the optical detection image to determine whether they are extended radio objects and confirm by eye if they are truly radio objects with spectroscopic counterparts. With this method, we added 18 radio galaxies with spectroscopically-confirmed redshifts to the radio confirmed sample. The numbers of extra radio galaxies confirmed in each of the LSS are listed in Table \ref{tab:matches}.

To account for radio doubles, i.e., those galaxies for which two distinct radio objects are physically associated with an individual optical object, as in the case of Fanaroff-Riley type I and II sources \citep{Fanaroff1974},  we used $\sigma_j = 5\arcsec$ as the search radius and ran the optical matching again using the spectroscopically-confirmed members. We examined the additional matches by overlaying radio flux contours for each match on the optical detection image to determine and confirm by eye if they are truly radio doubles. We found one radio double in SC1324. The radio flux density of this source is the combined flux density of two radio sources. 

In order to confirm our spectroscopic sample is not biased, we performed the following steps. We use $z_{phot}$ for radio-phot matches, when the photometric counterparts do not have a secure $z_{spec}$, to select potential LSS members. The radio-phot sample includes both these galaxies and the radio confirmed galaxies that are matched in the same radius ($\sigma_j = 1\arcsec$). This sample is described in more detail in Section~\ref{sec:photoobs} and Appendix~\ref{app:spec-photo}, named as radio-phot sample. The number of radio-phot sample objects in each LSS are listed in Table~\ref{tab:matches}.

When a radio source matches to a spectroscopic target with a blended serendip, we were required to determine the real spectroscopic counterpart. There are eight such cases across the five LSS: six in SC1604, one in SG0023, and one in RXJ1821. We determined the spectroscopic counterpart of a given radio source based on two methods. First, the spatial separation between the radio source and spectroscopic target and its serendip based on a careful de-blending of the object when possible: the further the optical object from the radio object, the less likely it will be matched to the radio object. In cases where this de-blending was not convincing, we examined the two-dimension spectrum of the blended sources and chose the brighter of the traced emission, since the probability of a given radio object matched to the brighter optical object is higher. We determined that the primary spectroscopic targets are the correct matches to all eight radio objects.

\section{Two-stage Radio Classification}
\label{sec:classification}
Radio emission can have an origin from synchrotron radiation and free-free radiation, both of which are active processes for a SFG. Conversely, the radio emission in AGN is
dominated by synchrotron and related processes. Therefore, we need a method to distinguish between radio emission due to star formation and that due to AGN activity. We define a ``hybrid galaxy"  (hereafter Hybrid) as a radio galaxy with emission potentially associated with both SF and AGN processes. 

The FIR and radio luminosities of star-forming galaxies are found to be tightly related, known as FIR/radio correlation (FRC; e.g~\citealp{dejong1985, Helou1985, Helou1988, Condon1992, Yun2001}). The total infrared ($8\sim1000 \mu m$, TIR)/radio luminosity ratio $q_{TIR}$, one indicator of the FRC, has been widely used to disentangle AGNs and SF galaxies (e.g.,~\citealp{Helou1985, Bell2003, Sargent2010, Padovani2009}), since the FIR emission at 250-1000$\mu m$ comes from optically thick regions of an intensely star-forming galaxy~\citep{Kennicutt1998}. Furthermore, the FRC has been found to be constant at z < 2~\citep{Sargent2010}, though, more recently, studies indicated a mild evolution on the slope of $(1+z)^{\gamma}$~\citep{Mao2011, Bourne2011, Schleicher2013}, where $\gamma = -0.12\pm0.04$~\citep{Magnelli2015}. 
We lack mid- or far-infrared observation in some of our cluster fields to derive TIR and thus are unable to adopt this method for the full sample. As a solution to this problem, we utilize MIPS 24$\mu$m observations from SC1604, SG0023 and RXJ1821, to derive $q_{TIR}$, and to develop a two-stage classification that relies on $L_{1.4GHz}$ and colour-Stellar mass normalized Radio Luminosity density(colour-SRL), using the $M(U)-M(B)_{rest-frame}$ as colour versus $L_{1.4GHz}$ modulated by the galaxy stellar mass. We obtain 107 radio galaxies from radio-spec matches at all redshifts which have 24$\mu$m detections at $\geq 3 \sigma$ (see Section~\ref{sec:photoobs}), noting that these galaxies are not necessarily confirmed to the LSSs. 

\subsection{Radio Luminosity Criteria}
We set the primary criteria to be that radio galaxies with $log(L_{1.4GHz}) > 23.8$ are classified as AGNs. 
Studies have found that almost all galaxies at $z \sim 1$ with radio power above this threshold are AGN, and is consistent with the maximal radio output of normal galaxies~\citep{Condon1992, Kauffmann2008, DelMoro2013}, and also consistent with the selection used by other studies (e.g.,~\citealp{Hickox2009}). Applying this criteria, we select 20 AGNs from the radio confirmed sample (see Table~\ref{tab:matches}). 
We note that the radio luminosity criteria means that all SFGs and Hybrids have luminosities below $10^{23.8}$; however, because of the second colour-SRL criteria (see Section \ref{sec:colour_SRL}), AGN can have radio luminosities less than $10^{23.8}$ as well.

\subsection{colour-SRL Criteria}
\label{sec:colour_SRL}
The second stage of our classification is the colour-SRL method. The dust extinction correction applied to the rest-frame $M_{U}-M_{B}$ colour follows the method in~\citet{Calzetti2000} with Av, derived from photometric catalogs, as the colour excess of the stellar continuum $Es(B - V)$ for each spectroscopic object. 
Studies have shown that AGN and star forming galaxies occupy different areas in the 4000\AA\ break D$_n$(4000) vs. radio luminosity normalized by stellar mass diagram for radio galaxies up to $z < 1.3$ \citep{Best2005,Smolcic2008}. The Pearson correlation coefficient of D$_n$(4000) and $M_U-M_B$ (dust corrected) restframe colour is $\sim 0.65$, using spectroscopically-confirmed galaxies which have \textbf{i)} a reliable measurement of D$_n$(4000), \textbf{ii)} a 3$\sigma$ detection in MIPS, and \textbf{iii)} which were matched to a significant ($>4\sigma$) radio source. Because of this strong correlation and our inability to measure D$_n$(4000) for every radio galaxy, we use the rest-frame $M_{U}-M_{B}$ colour instead. 

We calibrate the colour-SRL separation with $q_{TIR}$ classification. We utilize MIPS mid-infrared observations from SC1604, which is the only field with deep, wide-field MIPS observation amongst the five fields, and two narrow-field ($7' \times 7'$) observations from SG0023 and RXJ1821. To maximize our sample population, we take all radio sources with high quality spectroscopic counterparts in the entire field which also have $3\sigma$ 24$\mu$m detections. We obtain 107 galaxies from which to obtain our classification rules for the colour-SRL method via $q_{TIR}$ equation adopted from \citet{Lemaux2014}:
\begin{equation}
    	q_{TIR} = log(\frac{3.939\times 10^{26} L_{TIR}}{3.75 \times 10^{12} W}) - log(\frac{L_{1.4GHz}}{W Hz^{-1}}).
	\label{eq:qTIR}
\end{equation}

The main contribution to the uncertainty on $q_{TIR}$ comes from the spectra index ($\alpha$), which is typically in the range of $0.4 \le \alpha \le 1.0$ for 80\% of all types of galaxies in the redshift range $0 < z < 4$~\citep{Lemaux2014}, with $q_{TIR}$ to be affected up to 20\%. However, none of our results are sensitively changed at this level.

\begin{figure}
    \includegraphics[width=\columnwidth]{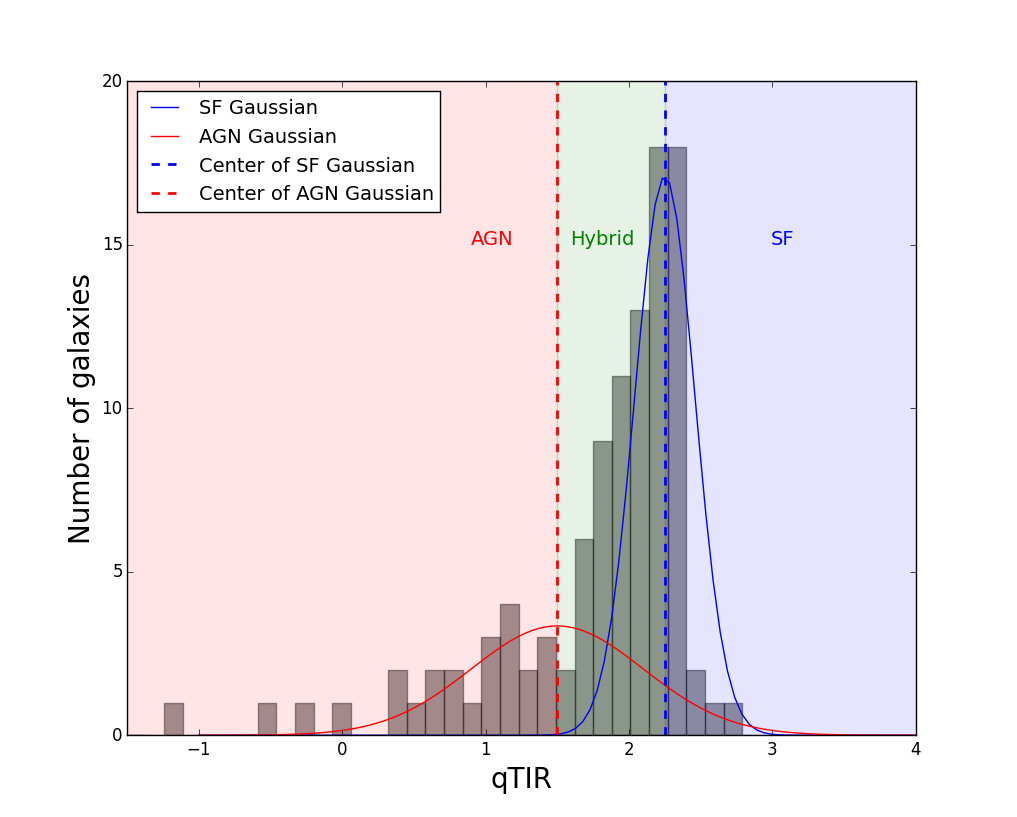}
    \caption{$q_{TIR}$ histogram. Including 107 radio objects with spectroscopic and mid-infrared counterparts. Fitting by a Double-Gaussian model, plotted in coloured solid line, centred at $q_{TIR} = 1.50$ $(\sigma = 0.60)$ and $q_{TIR} = 2.25$ $(\sigma = 0.21)$, marked as the vertical dashed lines. The red, green and blue regions are designated to be AGN, Hybrid, and SFG classified as qTIR.}
    \label{fig:qTIR}
\end{figure}

By fitting a double Gaussian model to the $q_{TIR}$ histogram shown in Figure~\ref{fig:qTIR}, we are able to separate radio galaxies into AGN dominated and star-formation dominated regions. Radio galaxies with the higher $q_{TIR}$ Gaussian fitting, centred at $q_{TIR} = 2.25$ $(\sigma = 0.21)$, represent the SF dominated population, while the Gaussian fitting with lower $q_{TIR}$, centred at 1.50 $(\sigma = 0.60)$ includes the AGN dominated population. 
Our star-formation Gaussian fitting agrees with $<q_{TIR}> = 2.16 \pm 0.08$ at $z \sim 1$, using a complete stellar-mass-selected ($M_* \ge 10^{10}~M_\odot$) sample of star-forming galaxies at $0 < z < 2.3$~\citep{Magnelli2015}.
Since there is a wide overlap between the AGN and SF regions, we choose AGN to have $q_{TIR} < 1.50$ (the centre of the lower Gaussian), SFGs to have $q_{TIR} \geq 2.25$, and Hybrids to populate the intermediate region of $1.50< q_{TIR} < 2.25$. 

Plotting these 107 radio galaxies in the rest-frame $M_{U}-M_{B}$ versus $L_{1.4GHz}/M_*(M_\odot)$ (see Figure~\ref{fig:UBall}), we found a clear separation between the three radio populations classified by $q_{TIR}$. Therefore, we set the separation lines, marked as dashed lines in Figure~\ref{fig:UBall}, for further use as a classification criteria in the full radio sample. If $q_{TIR}$ serves as the ground truth for the radio classification, we could test the relevance of our algorithm by adopting precision and recall\footnote{The precision is the fraction of the number of sources being selected in a given sub-class in both $q_{TIR}$ and colour-SRL over the number of sources selected using colour-SRL for that same sub-class. Recall is the fraction of the same numerator in the precision over the number of sources selected using $q_{TIR}$ for that same sub-class.}. 

\subsection{Result of AGN/Hybrid/SFG Classification and Bias}
\begin{figure}
  \includegraphics[width=\columnwidth]{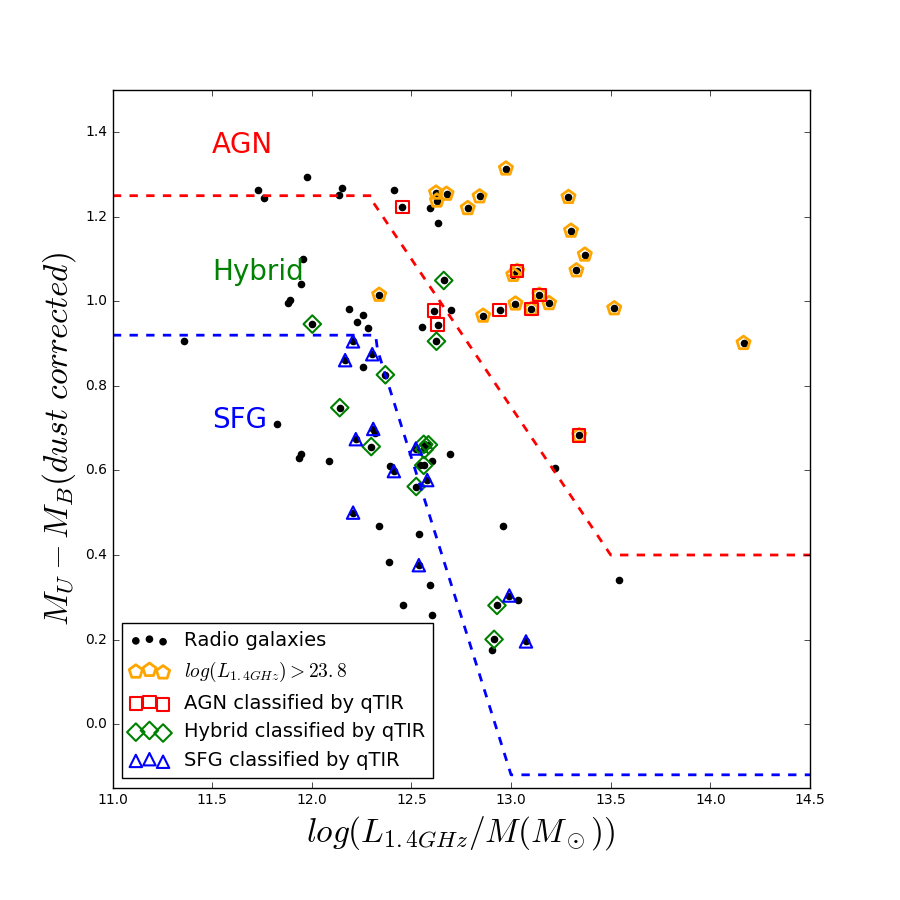}
    \caption{The rest-frame $M_{U}-M_{B}$ (corrected for dust extinction) as a function of the stellar mass normalized 1.4 GHz luminosity density for all radio confirmed galaxies (black dots) in the five LSSs. The red and blue dashed lines are used to seperate various radio sub-classes in the colour-SRL scheme. Dots with orange open pentagons are radio galaxies with $log(L_{1.4GHz}) > 23.8$. Dots with open markers (squares, triangles, diamonds) are coloured by their $q_{TIR}$ classification method, and are used set the colour-SRL criteria.}
    \label{fig:UBall}
\end{figure}

Using the above two-stage classification, we classified our radio galaxy sample into three subsamples: AGN, Hybrid, and SFG. This classification is intended to identify the process dominating their radio emission and does not necessarily exclude the existence of other sub-dominant components. 
We show the classification result in a colour-SRL diagram in Figure~\ref{fig:UBall}. The number of each sub-class are summarized in Table~\ref{tab:type}. 
We notice that 95\% of AGN selected in the first stage with $log(L_{1.4GHz}) > 23.8$ are classified as AGN in the second stage as well. Thus, our results are largely unaffected if we instead drop the radio luminosity criteria for AGN classification. 
We found that a few SFGs and Hybrids are present at large value of mass weighted radio luminosity, values as large as many of AGN. These large values are, however, the result of these galaxies having lower stellar mass values than their radio AGN counterparts. 
We should note that spectroscopic incompleteness might affect our radio sample, specifically we are likely underestimating the percentage of SFGs in the radio sample. However this will not affect our conclusions as we will discuss in Section~\ref{sec:properties}. 

\begin{table}
	\centering
	\caption{Number of AGN, Hybrids and SFGs in the radio confirmed sample in each LSSs}
	\label{tab:type}
	\begin{threeparttable}
	\begin{tabular}{lcccc} 
		\hline
		\hline
		Field & Total & AGN &  Hybrids &  SFGs \\
		\hline
		SC1604 & 41 & 14 & 17 & 10\\
		SG0023 & 8 & 2 & 3 & 3\\
		SC1324 & 25 & 12 & 6 & 8\\
		RXJ1757& 3 & 1 & 1 & 1\\ 
		RXJ1821 & 12 & 4 & 5 & 3\\
		\hline
		Total & 90 & 33 & 32 & 25\\
		\hline
	\end{tabular}
	 \end{threeparttable}
\end{table}

\subsection{X-ray Cross Match}
Cross-matching radio galaxies with X-ray confirmed AGNs, drawn from~\citet{Rumbaugh2016}, we found a total of five out of 89 radio galaxies have X-ray counterparts (see Table~\ref{tab:x-ray}). Three of these are classified as AGN in our two-stage radio classification scheme, consistent with AGN emitting at both wavelength regimes and in line with other studies (e.g.,~\citealp{Hickox2009, Lemaux2014}). Interestingly, one of the X-ray AGN is classified as Hybrid and another one as SFG. This may be explained by the host galaxy having X-ray bright in nucleus activity, while the radio emission is from concurrent star-formation activity. It is also possible that X-ray dominated in one stage of the AGN evolution, while radio dominated in a later stage~\citep{Hickox2009}. 

\begin{table*}
	\centering
	\caption{Radio and X-ray Crossmatch}
	\label{tab:x-ray}
	\begin{threeparttable}
	    \begin{tabular}{lccccccccc} 
		    \hline
		    \hline
		    Field & Radio & RA & Dec & Radio & X-ray & X-ray Full & $log(\eta)\tnote{2}$ & colour & Type\tnote{4} \\
		    & ID & & & Power & ID\tnote{1} & Luminoisty\tnote{1} &  & offset\tnote{3} & \\
		    \hline
		    SC1604 & 502 & 241.0647 & 43.1713 & 23.28 & 0 & 6.269E+43 &  0.871 &	0.038 & Hybrid\\
		    SC1604 & 568 & 241.1076 & 43.2126 & 23.67 & 2 & 2.111E+43 & 0.584 & 0.160 & AGN \\
		    SC1604& 681 & 241.1567 & 43.1494 & 24.30 & 6 & 6.175E+42 & 0.871 & 0.038 & AGN \\
		    RXJ1821 & 198 & 275.2819 & 68.3941 & 23.72 & 2 & 6.240E+42 & -0.418 & 0.336 & AGN \\
		    RXJ1821 & 237 & 275.3494 &  68.4424 & 23.15 & 0 & 7.728E+42 & 0.215 &	0.375 & SFG \\
	        \hline
	    \end{tabular}
	    \begin{tablenotes}
            \item[1] X-ray ID and Full band Luminosity adopted from \citet{Rumbaugh2016}.
            \item[2] $log(\eta)$ parameter is described in Section~\ref{sec:spatial}.
            \item[3] colouroffset parameter is explained in Section~\ref{sec:cc}.
            \item[4] The type of radio galaxy is based on our two-stage classification scheme.
        \end{tablenotes}
    \end{threeparttable}
\end{table*}

\section{Radio Galaxy and Host Properties}
\label{sec:properties}
With these classifications in place, we are now able to begin a deep analysis of different populations of radio galaxies to compare the properties of AGN, Hybrids, and SFGs. We have combined all radio galaxies in the five LSS for a complete, statistical analysis of radio galaxies across a variety of environments in the redshift range of $0.65 \le z \le 0.96$. With a relatively small redshift range, we choose to ignore redshift-driven evolutionary effects. In this section we show the preferences of the three sub-classes of radio host galaxies in terms of optical colour, stellar mass, radio luminosity, spectral properties, and environments. We summarize median values of these properties for the three radio populations in Table \ref{tab:median} and present the results of K-S tests comparing each pair of populations for all properties in Table \ref{tab:ks}. We find that the host galaxies are significantly different among these three radio galaxy populations.

\begin{table*}
	\centering
	\caption{Summary of Average Properties of the Three Radio Populations}
	\label{tab:median}
	\begin{threeparttable}
	    \begin{tabular}{lccccccccc} 
		    \hline
		    \hline
		    Radio  & $\langle Colour \rangle$\tnote{1} & Colour & $log(\langle M_*/M_\odot\rangle)$\tnote{3} & $log(\langle \eta \rangle)$\tnote{4} & $log(\langle 1 + \delta_{gal} \rangle)$\tnote{5} & $log(\langle L_{1.4GHz}\rangle)$\\
		    Type &  & Offset\tnote{2} &  &  & & \\
		    \hline
		    AGN & 4.62 & 0.10 & 11.16 $\pm$ 0.03 &  -0.74 & 0.67 & 24.01 $\pm$ 0.05\\
		    Hybrid & 3.38 & -0.74 & 10.74 $\pm$ 0.07 & -0.55 & 0.58 & 23.21 $\pm$ 0.02\\
		    SFG & 3.36 & -0.71 & 10.95 $\pm$ 0.03 & -0.51 & 0.48  & 23.21 $\pm$ 0.02 \\
	        \hline
	    \end{tabular}
	    \begin{tablenotes}
            \item[1] Median value of Rest-frame $M_{NUV} - M_{r}$;
            \item[2] Median value of colour offset parameter, derived in Section~\ref{sec:cc};
            \item[3] With standard error of the median listed;
            \item[4] $\eta = R_{proj}/R_{200} \times |\Delta_\nu|/\sigma_\nu$, see explanation in Section~\ref{sec:spatial};
            \item[5] See explanation on $log(1 + \delta_{gal})$ in Section~\ref{sec:spatial}.
        \end{tablenotes}
    \end{threeparttable}
\end{table*}

\begin{table*}
	\centering
	\caption{Summary of p-values from K-S test}
	\label{tab:ks}
	\begin{threeparttable}
	    \begin{tabular}{lccccccccc} 
		    \hline
		    \hline
		    Sample-Sample\tnote{1}  & Colour\tnote{2} & Colour & $log(\langle M_*/M_\odot\rangle)$ & $log(\langle \eta \rangle)$\tnote{4} & $log(\langle 1 + \delta_{gal} \rangle)$\tnote{5} & $log(\langle L_{1.4GHz}\rangle)$\\
		    Type &  & Offset\tnote{3} &  &  &  & \\
		    \hline
		    AGN-Hybrid & $10^{-5}$ & $10^{-6}$ & $10^{-3}$ & 0.44 & 0.10 & $10^{-11}$\\
		    AGN-SFG & $10^{-8}$ & $10^{-6}$  & 0.01 & 0.10 & 0.24 & $10^{-9}$ \\
		    Hybrid-SFG & 0.18 & 0.09 & 0.05 & 0.66 & 0.91 & 0.93\\
            Hybrid-Mixed\tnote{6} & $10^{-6}$ & 0.02& $10^{-4}$ & 0.09 & 0.08 &$10^{-4}$ \\
            Radio-Spec & 0.004 & 0.006 & $10^{-14}$& 0.49 & 0.15 & --\\
	        \hline
	    \end{tabular}
	    \begin{tablenotes}
        	\item[1] Two compared galaxy samples. 
            \item[2] Rest-frame $M_{NUV} - M_{r}$;
            \item[3] Colour offset parameter, derived in Section~\ref{sec:cc};
            \item[4] $\eta = R_{proj}/R_{200} \times |\Delta_\nu|/\sigma_\nu$, see explanation in Section~\ref{sec:spatial};
            \item[5] See explanation on $log(1 + \delta_{gal})$ in Section~\ref{sec:spatial}.
            \item[6] A sample of mixture of AGN and SFGs, drawn with the same number of galaxies in the Hybrid population from AGN and SFG population for 100 trials, allowing the relative number of AGN and SFG to vary in each draw. 
        \end{tablenotes}
    \end{threeparttable}
\end{table*}

\subsection{Radio Galaxy colours}\label{sec:cc}

\begin{figure}
    \includegraphics[width=\columnwidth]{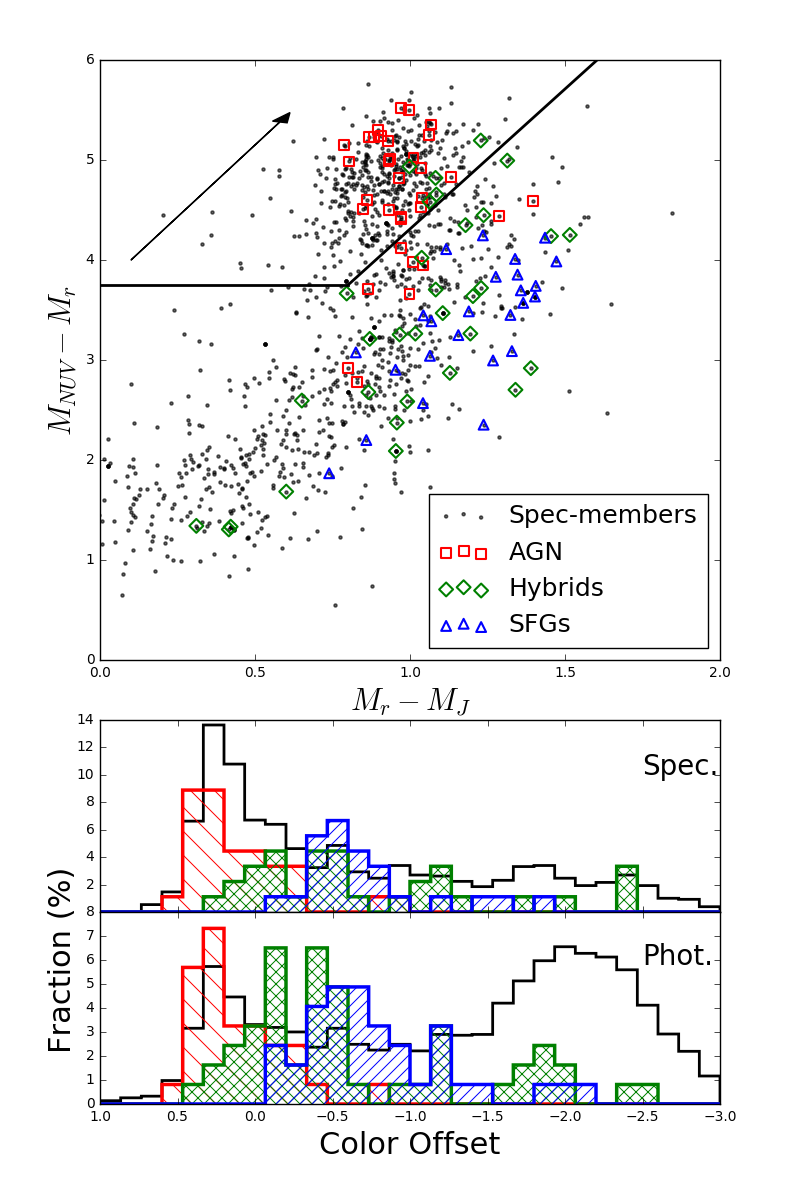}
    \caption{\textit{Top}: The rest-frame colour-colour diagram ($M_{NUV} - M_{r}$ versus $M_{r} - M_{J}$). Dots are spectroscopically-confirmed members in the five LSSs. Radio galaxies are marked with open squares, coloured by their type. The solid lines show the adopted divisions between the star-forming and quiescent galaxy populations. The left upper region is the quiescent region, and the right lower region is the star-forming region. The arrow represents the movement in the phase space when a stellar continuum extinction of Es(B - V) = 0.3 is applied~\citep{Calzetti2000}. \textit{Bottom 1}: colour offset histogram of spectroscopically-confirmed members (black line, scaled by the total number of spectroscopic members), and radio confirmed galaxies (hatched colour histogram, scaled by the number of radio confirmed galaxies). \textit{Bottom 2}: colour offset histogram of photometric members (black line, scaled by the total number of photometric members), and radio-phot sample (hatched colour histogram, scaled by the number of radio-phot sample). Positive values represent objects in the quiescent region.}
    \label{fig:cc}
\end{figure}

We use a two-colour selection technique proposed by~\citet{Williams2009} to divide the galaxies into two catagories: quiescent and star-forming (SF). We adopt the rest-frame of $M_{NUV} - M_{r}$ versus $M_{r} - M_{J}$ colour-colour diagram following separation lines from~\citet{Lemaux2014}, where galaxies with $M_{NUV} - M_r > 2.8(M_r - M_J) + 1.51$ and $M_{NUV} - M_r > 3.75$ are considered quiescent. We show the colour-colour diagram for spectroscopically-confirmed members (black dots) and radio galaxies, marked with colour open squares depending on their radio type, in Figure~\ref{fig:cc} top panel. The advantage of colour-colour diagrams, rather than colour-magnitude, is that dusty star-forming galaxies move to the right top along a diagonal axis, instead of mixing with quiescent galaxies. This is a typical limitation when performing analysis solely with a colour-magnitude diagram.

We generate a colour offset histogram in the bottom panel of Figure~\ref{fig:cc}, with the upper of the two histograms for the radio confirmed members and the lower for the radio-phot sample, where ``colour offset" is the perpendicular offset from quiescent and star-forming separation line, with positive representing galaxy in the quiescent region. With the help of the colour offset, we are able to quantitatively examine the distribution of radio sub-classes located in the 2D colour-colour diagram. In the radio confirmed members colour offset histogram, we see a separation between the population of AGN host galaxies and SFG host galaxies. 72\% of the AGN sample has colour offset above zero (i.e., in the quiescent region). A small fraction of the AGN inhabiting the star-forming region, and most of them are close to the dividing line (colour offset $< -0.5$). AGN are found to be hosted largely by red, quiescent galaxies, as we expect from our second selection criteria, although the radio luminosity criteria does not limit radio AGN hosts to be only red galaxies. In addition, we note that redder in dust-corrected $M_U-M_B$ does not necessarily require galaxies to be classified as quiescent in the $M_{NUV}-M_r/M_r-M_J$ colour-colour diagram, as evidenced by the large number of Hybrid galaxies which are situated in the star forming locus of this diagram. 
SFGs are narrowly displayed in the dusty active region, with none of the SFGs located in the quiescent region confirming that SFGs in radio samples represent a dusty SFG population. The colour offset histogram of Hybrid host galaxies shows that most are located in the star-forming region and extends to lower values than that of SFGs.

We employed the Kolmogorov-Smirnov statistic (K-S) test on the AGN, Hybrid and SFG samples to test whether any two are consistent with being draw from the same distribution. The two tailed p-value result of the K-S test gives the probability that two distributions are drawn from the same distribution. In this paper, we adopt that if the p-value > 0.1, we cannot reject the hypothesis that the two distributions are drawn from the same distribution. Otherwise, we say the probability of drawing from the same distribution is very small. We note that none of the results will change meaningfully if we use p-value = 0.05 as the threshold. The p-value between the colour offset distribution of AGNs and SFGs (or Hybrids) is $\sim 10^{-6}$, which confirms that AGN host galaxies are different from Hybrids and SFGs. The host galaxies of SFG and those of Hybrid populations are similar, with p-value of $\sim 0.1$. Therefore, here, we are not able to confirm a difference between Hybrids and SFGs. 

We also tested whether the Hybrid population could be a mixture of AGN and SFGs. We drew 100 samples having the same number of galaxies as were in the Hybrid population from AGN and SFG population, allowing the relative number of AGN and SFG to vary in each draw. We ran the K-S test for each sample and returned the mode of p-values(see Appendix~\ref{app:control} for the definition of the mode of the p-values). The resultant value was 0.02, strongly suggesting that the Hybrid population is not a mixture of AGN and SFGs, the one exception being the few cases where SFGs dominated the comparison sample.

However, if we want to compare the percentage of each radio sub-class of the total radio sample, we may be biased by the incompleteness of our spectroscopic survey, since, according to the $M_{NUV} - M_{r}$ colour, the spectroscopic sample becomes unrepresentitive when $M_{NUV} - M_{r} \le 2.5$ (see more detail on testing the spectroscopic representativeness in Appendix~\ref{app:spec-photo}). In the lower of the two histograms in the bottom panel of Figure~\ref{fig:cc}, we plot the colour offset histogram for photometric members and the three sub-classes from the radio-phot sample, using the same classification scheme (AGN-phot, Hybrid-phot, and SFG-phot). As shown in Figure~\ref{fig:cc}, there is a bimodal distribution for photometric members, however the star-forming peak is absent for the spectroscopically-confirmed members. This is consistent with our spectroscopic sample not being representative of all galaxies at bluer colours. As for radio galaxies, although adding the $z_{phot}$ selected radio galaxies increases the total radio sample, the coverage and distribution of each radio sub-population in the colour offset does not change. This indicates that spectroscopic incompleteness does not bias our results for different types of radio galaxies. Hence, we confirm that AGN are red and quiescent galaxies, SFGs are bluer, active star-forming galaxies, and Hybrids are dispersed in colour, though with a slight preference to lie in the star-forming region. 

\subsection{Stellar Mass}\label{sec:sm}

\begin{figure}
    \includegraphics[width=\columnwidth]{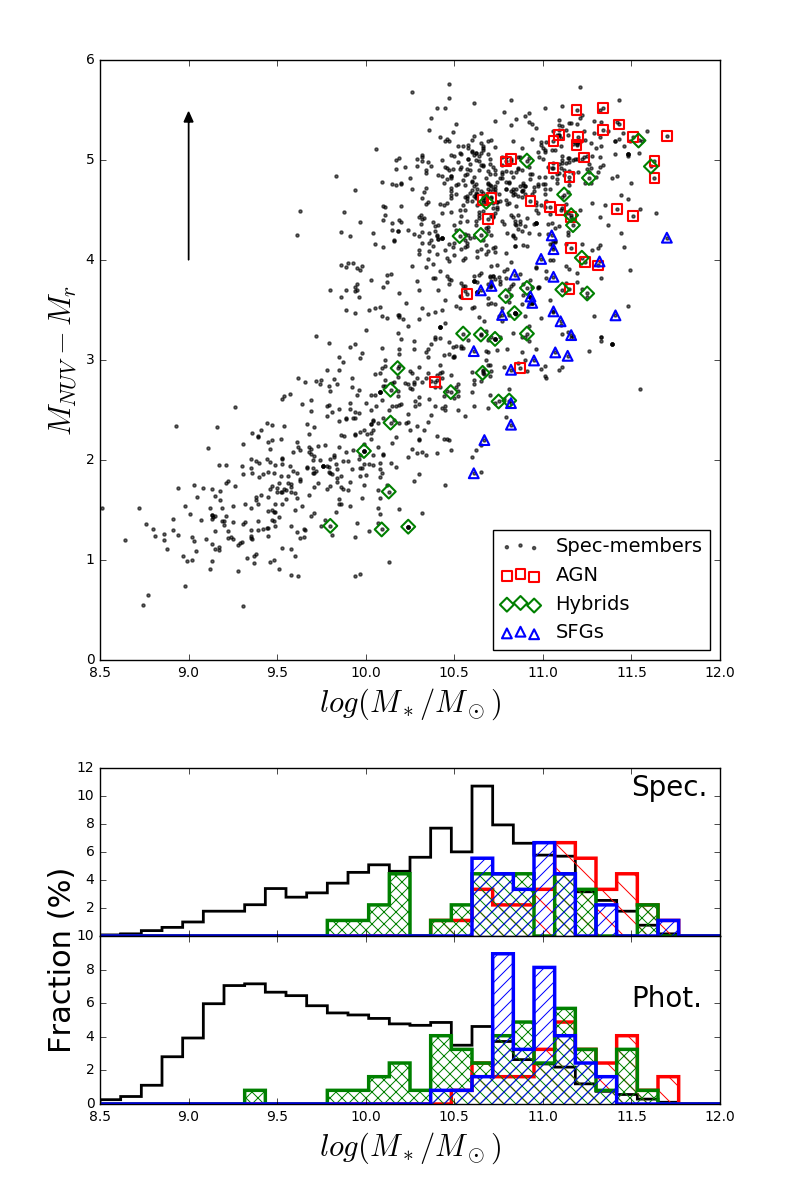}
    \caption{\textit{Top}: The rest-frame of $M_{NUV} - M_{r}$ versus $log(M_*/M_\odot)$. Dots are spectroscopically-confirmed members in the five LSSs. Radio galaxies are marked with open squares, coloured by their type.  The arrow represents the movement in the phase space when a stellar continuum extinction of Es(B - V) = 0.3 is applied~\citep{Calzetti2000}. \textit{Bottom 1}: Stellar mass histogram. Black histogram represent spectroscopically-confirmed members scaled by the total number of members. Hatched colour histograms are the three radio populations scaled by the total number in the radio confirmed sample. \textit{Bottom 2}: Stellar mass histogram of photometric members and radio-phot. Black histogram represent photometric members scaled by the total number of photometric members. Hatched colour histograms are the three radio-phot populations scaled by the total number in the radio-phot sample.}
        \label{fig:mass}
\end{figure}

The phenomenon of ``mass quenching" results in galaxies with higher stellar masses being observed with higher quiescent fractions (e.g.~\citealp{Brinchmann2000, Kauffmann2003a, Bell2005, Peng2010, Cucciati2016}). This result indicates that the most massive galaxies have already been quenched before $z\sim1$~\citep{Ilbert2013, Darvish2016}. The origin of this mass-dependent SF cessation or quenching is still unclear. It could be the effect of feedback from AGN~\citep{Hopkins2006, Somerville2008}. 
A compelling explanation is the heating of accreting cold gas as it falls into massive halos~\citep{Birnboim2003}. Studies at z < 1 have found that the majority of radio-AGN are located in high-stellar mass quiescent galaxies~\citep{Best2005, Best2007, Kauffmann2008}. Therefore, it is necessary to study the distribution of stellar mass for different types of radio galaxies. Later, we will control this mass-dependent quenching effect by matching radio sub-classes in colour and stellar mass to non-radio galaxies (see Section~\ref{sec:control}). 

In the top panel of Figure~\ref{fig:mass}, we plot the rest-frame of $M_{NUV} - M_{r}$ versus $log(M_*/M_\odot)$ for the three radio sub-classes and spectroscopically-confirmed members. We see that radio host galaxies appear at the higher stellar mass end of the entire LSS member population. This preference for imassive hosts is seen more clearly in the stellar mass histogram, shown in the upper of the two histogram in the bottom panel of Figure~\ref{fig:mass}. 
The average stellar mass of radio confirmed members is $\langle M_*\rangle = 10^{11.0}~M_\odot$, compared to $\langle M_*\rangle = 10^{10.4}~M_\odot$ for the full LSS spectroscopically-confirmed members ($\langle M_*\rangle = 10^{10.7}~M_\odot$ if the stellar mass cutoff is included, see Appendix~\ref{app:spec-photo} for explanation). This result is consistent with studies showing that the probability for a galaxy to be a radio source increases with increasing stellar mass (e.g., \citealp{Ledlow1996}). We employ the K-S test to verify our hypotheses that the distribution of the radio confirmed galaxies differs from that of the spectroscopically-confirmed members and calculate a p-value of $\sim 10^{-14}$. With such a small p-value, we conclude that the stellar mass distribution of the radio confirmed galaxy sample is different from that of the spectroscopically-confirmed members. 

We find that the AGN host galaxy sample skews toward higher stellar masses among the three radio sub-classes, while Hybrids evenly extend from $M_* \sim 10^{9.8}~M_\odot$ to $M_* \sim 10^{11.8}~M_\odot$, even lower than the SFG hosts. On the other hand, the population of SFG shows a narrow peak centred at $M_* \sim 10^{11}~M_\odot$. We conclude that none of the three sub-classes is drawn from the same distribution, a conclusion confirmed by K-S test. Again, we confirm that the Hybrid population is not a mixture of AGN and SFGs, using the same sampling algorithm as in the colour offset analysis (Section~\ref{sec:cc}) with a small p-value mode ($\sim 10^{-4}$).

We then examined photometric members and radio-phot in stellar mass. We show the stellar mass histogram of photometric members and radio-phot sample in the lower of the two histograms in the bottom panel of Figure~\ref{fig:mass}. Although the distribution of photometric members is different from the distribution of spectroscopic members, the conclusion about radio galaxies and each sub-class are likely not affected.
We notice that we do not observe radio galaxies with stellar mass below $10^{9.8}~M_\odot$ in the radio confirmed member sample, with only one in the radio-phot sample. There are two possible reasons for this lack of low stellar mass radio galaxies: insufficient radio depth or a trigger mechanism that is correlated with $M_*$. 
We calculate the stellar mass limit of the SFGs by using the SFR-$M_*$ relation~\citep{Tomczak2016}. The lowest radio luminosity for our SFGs is $log(L_{1.4GHz, min}) = 22.96$, corresponding to $SFR_{min} = 30.23~M_\odot yr^{-1}$, using the star formation rate formula from 1.4GHz from~\citet{Bell2003} and converting Salpter IMF to Charibier IMF by multiplying by a factor of 0.6. Assuming the SFGs lie on the main locus of the SFR-$M_*$ relation of $0.5 < z < 1.0$ from Figure 4 in~\citet{Tomczak2016}, we find, at such SFR, the stellar mass is $\sim10^{10.5}M_\odot$. Therefore, it appears that it is our radio depth which prevents us from observing galaxies below this stellar mass limit.
The issue is likely similar for hybrid galaxies. However, the tail of lower stellar mass hybrid galaxies is likely due to the fact that less SFR is needed to produce the radio power density by virtue of the additive presence of an AGN component. Therefore, these galaxies can be less stellar massive for a fixed radio depth survey. 
Regarding the radio AGN hosts, in a large and deep survey of radio AGN from $0 \le z \le 1.3$,~\citet{Smolcic2009a} found essentially no AGN hosts with stellar masses below $10^{10.7}~M_\odot$. It appears to be a stellar mass limit below which we do note see AGN dominating the radio emission. 
Thus, it is likely that this is a requisite stellar mass for radio AGN activity, and our lack of observing hosts with lower stellar mass is not due to issues related to the depth of our radio imaging. 

\subsection{Radio Luminosity}\label{sec:luminosity}

\begin{figure}
    \includegraphics[width=\columnwidth]{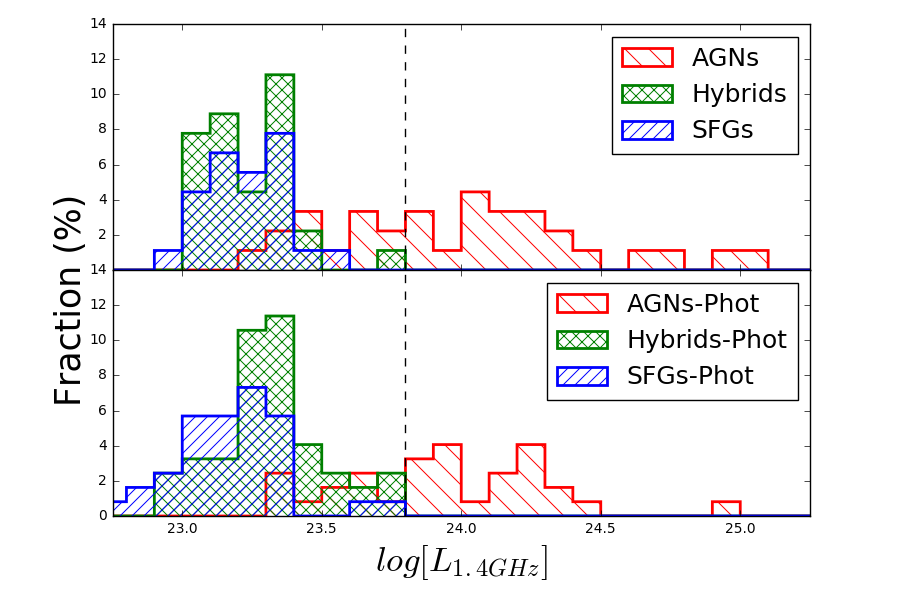}
    \caption{Radio Luminosity histogram of $log(L_{1.4GHz})$. The dashed black lines at $log(L_{1.4GHz}) = 23.8$ are used as the first criteria to separate AGN.~\textit{Top}: The radio Luminosity histogram of the three sub-classes AGN/Hybrid/SFG in the radio confirmed sample, scaled by the number of total radio confirmed galaxies.~\textit{Bottom}: The radio Luminosity histogram of AGN-, Hybrid- and SFG-phot sample, scaled by the total number objects in the radio-phot sample.}
    \label{fig:lum}
\end{figure}

Studies of AGN radio luminosity show a strong correlation with the black hole mass (MBH) and an anti-correlation with Eddington ratio  ($\lambda = L_{bol}/L_{Edd}$), using a variety of AGN classification methods (e.g.~\citealp{Laor2000, Ho2002, McLure2004, Sikora2007, Chiaberge2011, Sikora2013, Ishibashi2014}). On the other hand, thanks to the well established FRC~\citep{Kennicutt1998,Condon1992, Yun2001}, radio luminosity is a good indicator of the SFR for SFGs not modulated by the dust content of a galaxy and with a well-calibrated $SFR_{1.4GHz}$ formula~\citep{Bell2003, Hopkins2003}. However, the two populations have to be carefully separated to obtain their specific properties (e.g.~\citealp{Smolcic2008, Rees2016}). Therefore, it is necessary to study the radio luminosity of the three radio sub-classes separately. 

In Figure~\ref{fig:lum} top panel, we plot the histograms of $log(L_{1.4GHz})$ for the three radio sub-classes. We see the AGN population dominates the high radio luminosities, although part of this could be due to our primary classification criteria where galaxies with $L_{1.4GHz} > 10^{23.8}~W~Hz^{-1}$ are classified as AGN. However, as we noted earlier, nearly all such galaxies would have been classified as AGN in the $q_{TIR}$/colour-SRL classification. Radio luminosities of AGN extend over two orders of magnitude. Hybrids and SFGs are dominant in the region through $L_{1.4GHz} < 10^{23.5}~W~Hz^{-1}$. 
Our median values for AGN and SFGs are consistent with $log(L_{1.4GHz}^{AGN}) = 24.0$ and $log(L_{1.4GHz}^{SFG}) = 23.2$, respectively, for radio galaxies at $z \ge 1.3$~\citep{Smolcic2008}. 
We confirm the difference via the K-S test, where the luminosity distribution of the AGN differ significantly from these of the SFG and Hybrid classes. However, the SFGs and Hybrids appear consistent with being drawn from the same distribution at a confidence level of 93\%. Again we tested whether the Hybrid class is from a mixture of AGN and SFGs, using the same sampling algorithm used in Section~\ref{sec:cc} and Section~\ref{sec:sm}. The p-value mode is $\sim 10^{-4}$, which agrees with the results of the colour offset and stellar mass analyses that Hybrids could not be a mixture of AGN and SFGs. 

In the bottom panel of Figure~\ref{fig:lum}, we plot the histogram of log(L$_{1.4GHz})$ for the three sub-classes separated from the radio-phot sample. We see that the three histograms span the same as the three histograms in the top panel. 
Therefore, even adding photometric members in, none of the three radio luminosity distributions would change.

\subsection{Spectral Anaysis}
\label{sec:spectra}
The region where galaxies reside in the $M_{NUV} - M_{r}$ vs. $M_{r} - M_{J}$ phase space, combined with their stellar mass, could be indicative of the time since their star formation episode~\citep{Ilbert2010,Lemaux2014a, Moutard2016}. Such differences suggest that the AGNs are dominated by older stellar populations than SFGs, with Hybrid galaxies difficult to interpret. Fortunately, we have high signal-to-noise (S/N) DEIMOS spectra that contain several age sensitive features (e.g. $D_n(4000)$, $H\delta$ $\lambda4101${\AA}, G-band $\lambda4305${\AA}). These spectra, in conjunction with the broad band magnitudes, allow us to place much stronger constraint on internal extinction, as a result of breaking the degeneracy between the stellar age of a galaxy and its dust content (e.g.,~\citealp{Thomas2016}). We are then able to estimate ages more precisely. 

\begin{figure*}
	 \includegraphics[width=\textwidth]{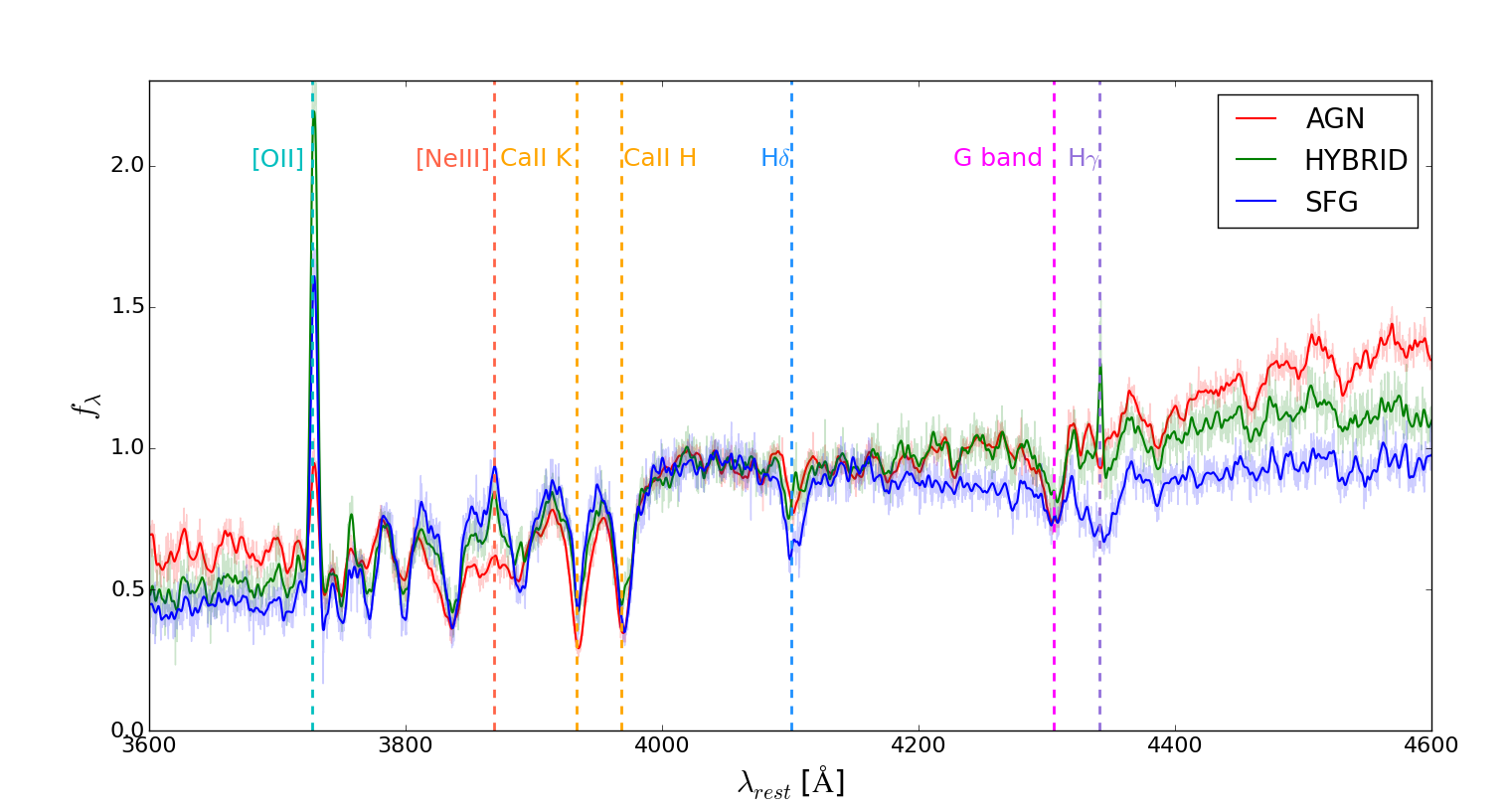}
	\caption{Inverse-variance, unit-weighted coadded DEIMOS spectra of the AGN (red), Hybrid (green) and SFG (blue) sub-classes. Important spectral features are indicated by vertical dashed lines and labeled.}
	\label{fig:spectra}
\end{figure*}

The DEIMOS spectra are combined (hereafter "coadded") within each sub-classes through an inverse variance-weighted average after shifting each individual spectrum to the rest frame, interpolating onto a standard grid with constant plate scale of $\Delta\lambda = 0.33/(1+z_{min})$ (where $z_{min}$ is the minimum $z_{spec}$ for each sample), and normalizing each spectrum to an average flux density of unity (e.g., unit weighted) following the methodology described in~\citet{Lemaux2012}. The coadded spectra of AGN, Hybrids and SFGs are shown in Figure~\ref{fig:spectra}. 

\begin{table*}
	\centering
	\caption{Composite Spectral Properties}
	\label{tab:spectra}
	\begin{threeparttable}
	    \begin{tabular}{lccccc} 
		    \hline
		    \hline
		    Sub-sample & Num. of & EW([\ion{O}{ii}]) & EW(H$\delta$) & $D_n(4000)$ & Age\tnote{2} \\
		    & galaxies\tnote{1} &  ({\AA}) & ({\AA}) & & (Gyr) \\
		    AGNs & 27 & -3.90$\pm$0.12 & 1.46$\pm$0.08 & 1.572$\pm$0.004 & $1.88^{+0.93}_{-0.55}$\\
		    Hybrids & 26 & -19.74$\pm$0.18 & 2.82$\pm$0.14 & 1.293$\pm$0.005 & $0.84^{+0.42}_{-0.31}$\\
		    SFGs & 24 & -11.52$\pm$0.15 & 4.79$\pm$0.12 & 1.203$\pm$0.003 & $0.24^{+0.16}_{-0.10}$\\
	        	    \hline
	    \end{tabular}
	    \begin{tablenotes}
	    	\item[1] The number of galaxy spectra in each sub-class that are used in the coadd. We exclude spectra from LRIS due to the poorer resolution, bad spectra due to reduction artifacts, and any type-1 AGN.
	    	\item[2] Time since last major star formation episode began.
             \end{tablenotes}
         \end{threeparttable}
\end{table*}
\begin{figure*}
	 \includegraphics[width=\textwidth]{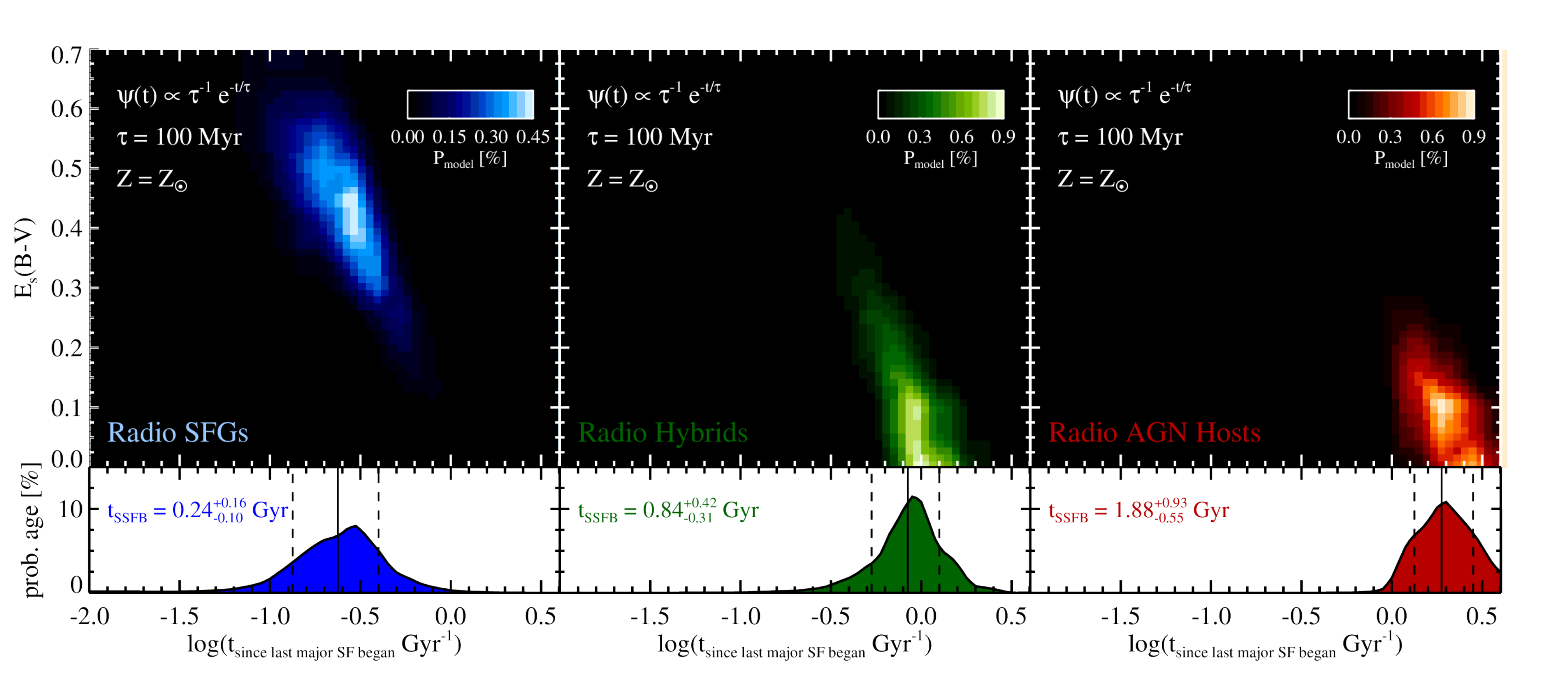}
	 \caption{Zoom in of the probability density maps (PDMs) for luminosity-weighted stellar ages and stellar extinctions of the average SFG (upper left panel), Hybrid (upper middle panel), and AGN (upper right panel) galaxies estimated from fitting stellar synthesis models simultaneously to the stacked DEIMOS spectra and stacked photometry. In both panels, 99.9$\%$ of the probability density of the full PDMs are contained within the displayed area. The form of the SFH, the e\-folding time of the exponential decline, and the stellar phase metallicity used for the fitting are given in the top of each panel. A scale bar is shown in the top right of each panel and maps the colours to their associated probabilities. The bottom panels show the extinction marginalized one-dimensional probability distribution functions (PDFs) of the luminosity$-$weighted stellar age for each sample. The median value of each PDF along with the associated effective 1$\sigma$ uncertainties is reported in the left of each panel.}
	 \label{fig:extinction}
\end{figure*}

The most outstanding difference between the coadded spectra is the strong [OII], $H\delta$ and $\lambda4305${\AA} emission of the average Hybrid galaxy. These features are less prominent or absent in the spectrum of the average SFG or AGN spectra. These characteristics emphasize that the Hybrid sub-class is a different population from a mixture of AGN and SFGs. The [OII] emission line is often used as a proxy of current star formation, especially at high redshift, when the H$\alpha$ emission line is shifted out of the optical~\citep{Poggianti1999}. However, [OII] emission can also be generated through low-ionization nuclear emission-line region galaxies (LINERs) and Seyfert processes~\citep{Yan2006, Lemaux2010, Kocevski2011}. 
The strength in the continuum break at 4000{\AA}, quantified as $D_n(4000)$~\citep{Balogh1999} is higher for AGN, indicating older galaxies, on average, compared to the Hybrid and SFG populations. 

Along with the coadded spectra, broad-band photometry were coadded following the method described in~\citet{Lemaux2016}. The coadded spectrum and photometry for each sample, after interpolating over emission features, was fitted simultaneously to synthetic models to investigate the evolutionary states of the three populations. In Figure~\ref{fig:extinction}, we show the probability density maps (PDMs) of $E_s(B-V)$ versus time since the last major star formation event began ($t_{SSFB}$), fitting to~\citet{Bruzual2003} models with the model parameters set to the values shown in each panel. Significant degeneracy exists between $t_{SSFB}$ and extinction for the SFG population (left panel). 
The high level of dust extinction in the SFG sub-class confirmed our earlier analysis, where we saw that SFGs are in the dusty star-forming region in colour-colour diagram (see Figure~\ref{fig:cc}). Meanwhile, the average age and dust extinction of the SFG sub-class indicates a starbursting phase.

Though the Hybrid population has more modest dust extinction on average, we do see a tail extending to higher extinction, while the AGN hosts exhibit the most modest dust content. In the bottom panel, we show the one-dimensional $t_{SSFB}$ PDF generated by adding probabilities of all values of $E_s(B-V)$ for each age step in each radio sub-classes. The median value of the PDF is marked by a solid line and $\pm 1\sigma$ values are denoted in dashed lines. The SFG population has a median $t_{SSFB} = 0.25~Gyr$, significantly younger than Hybrid and AGN populations. The measurements of EW[OII], EW[$H\delta$], $D_n(4000)$ and age are listed in Table~\ref{tab:spectra}. 

All analyses presented in this section support the conclusion that the AGN are at the oldest evolutionary stage among the three radio sub-classes. The host galaxies, on average, have ceased star formation $\sim 1.9~Gyr$ ago, consistent with the result in~\citet{Best2014} in which a time delay of 1.5-2 Gyr between the quenching of star formation and the onset of jet-mode radio-AGN activity was determined (also known as LERGs, as will be discussed in Section~\ref{sec:discussion}). 

On the other hand, perhaps the most intriguing observation is the difference in the spectral of the Hybrid population. Hybrids have a strong [\ion{O}{ii}], [\ion{Ne}{iii}], and H$\delta$ emission lines shown in the Figure~\ref{fig:spectra}. It is clear that a mixture of AGN and SFGs could have a composite spectra that on average, show the same continuum emission as the Hybrid composite spectra. However, based on Monte Carlo testing, we find that a combination sample of AGN and SFGs could not coadd in a way to exhibit such strong emission features. Therefore, we are confident to conclude that Hybrid galaxies are not some mixture of AGN and SFGs, but rather are a separate class powered by a different mechanism. Hybrids show evidence of, on average, a younger stellar population than AGN, strong ongoing star formation, and host galaxies with the lowest stellar masses and the farthest from the separation of the quiescent and the star forming region. This assertion will be quantified in Section~\ref{sec:hybrid}. 


\subsection{Spatial Distribution}
\label{sec:spatial}
\subsubsection{AGN/Hybrid/SFG Comparison}
\label{sec:radio_spatial}
We have shown in the previous sections the colour, stellar mass, radio luminosity and spectra for each radio sub-class. In this section, we explore the effect of environment, both global and local, on the three radio galaxy populations. The environment in which a galaxy resides plays an important role on its formation and evolution. 
In the local universe, radio AGN are preferentially found in the core of galaxy clusters, while star-forming galaxies are broadly distributed~\citep{Miller2002}. 
In terms of local environments, radio AGN also tend to be located in local overdense environments by examining local projected galaxy densities~\citep{Best2004} and real-space clustering properties up to redshifts $z \sim 0.5$ (e.g.~\citealp{Magliocchetti2004, Lindsay2014}). 
Therefore, it is imperative to explore the role of both the clusters/groups and local environment on the three radio sub-classes at $z \sim 1$.

We choose $R_{proj}/R_{200}$ versus $|\Delta\nu|/\sigma_\nu$~\citep{Carlberg1997, Balogh1999, Biviano2002, Haines2012, Noble2013, Noble2016} as the metric for global environment, as it probes the time-averaged galaxy density to which a galaxy has been exposed. 
$R_{proj}$ is the distance of a given galaxy to the nearest cluster centre, and $R_{200}$, the radius at which the matter density is 200 times the critical density. The centres of the clusters/groups are obtained from the i$^\prime$-luminosity-weighted centre of member galaxies calculated using the method described in~\citet{Ascaso2014}. $\Delta\nu$ is the velocity offset of the galaxy from the systemic velocity of the cluster/group, and $\sigma_\nu$ is the line of sight velocity dispersion of cluster/group member galaxies. The systemic velocity and the measured line-of-sight (LOS) galaxy velocity dispersion ($\sigma_\nu$) within the cluster/group are measured using the method of~\citet{Lemaux2012}. We define $\eta = R_{proj}/R_{200} \times |\Delta\nu|/\sigma_\nu$ (instead of p as in~\citealp{Noble2013} to distinguish it from the p-value in the K-S test and the probability threshold $p_i$), sometimes called caustic lines, to quantify the global environment.

In addition, we describe local overdensity, $log(1+\delta_{gal})$, derived from the Voronoi Monte-Carlo technique described in~\citet{Lemaux2016}. In the Voronoi Monte-Carlo technique, spectroscopically-confirmed members are sliced into various redshift bins. For each Monte-Carlo realization, photometric objects without a high quality $z_{spec}$ are sampled based on the $1\sigma$ uncertainty of the $P(z)$ (described in Section \ref{sec:photoobs}). In other words, a certain portion of photometric objects are added to create a new $z_{phot, MCi}$ for that realization, and Voronoi tessellation is performed on each realization of the redshift slice on all $z_{spec}$ and $z_{phot, MCi}$ that fall within that redshift bin. For each realization of each slice, a grid of $75\times75 h_{70}^{-1}$ proper kpc is created to sample the underlying local density distribution. The local density at each grid value for each realization and slice is equal to the inverse of the Voronoi cell area, and final local densities, $\Sigma_{VMC}$, are then computed by median combining the values of 100 realizations of the Voronoi maps for each grid point in each redshift slice. The local overdensity value for each grid point is then computed as $log(1 + \delta_{gal}) = log(1 + (\Sigma_{VMC} - \tilde{\Sigma}_{VMC})/\tilde{\Sigma}_{VMC})$, where $\tilde{\Sigma}_{VMC}$ is the median $\Sigma_{VMC}$ for all grid points over which the map is defined. We note that we adopt local overdensity rather than local density to mitigate issues of sample selection and differential bias on redshift. 

\paragraph{Global Environments}
\label{sec:global}

With $\eta$ defined by $R_{proj}/R_{200} \times |\Delta\nu|/\sigma_\nu$, it is useful to separate different galaxy populations associated with an individual group or cluster. Adopting definitions based on N-body simulations and semi-analytic simulations (see~\citealp{Noble2013}, reference therein), $|\eta| < 0.1$ is virialized core region. The intermediate region is $0.1 < |\eta| < 0.4$, and could be composed of at least some backsplash galaxies. Simulations suggested backsplash galaxies have been inside the viral radius at an earlier time and rebounded outward (e.g.,~\citealp{Balogh2000, Gill2005}). The outer region with $0.4 < |\eta| < 2$ should pick out galaxies that have been recently accreted, often known as the infall region~\citep{Haines2012}. Galaxies with $\eta > 2$ are likely not associated with any group or cluster. 

\begin{figure*}
    \includegraphics[width=\textwidth]{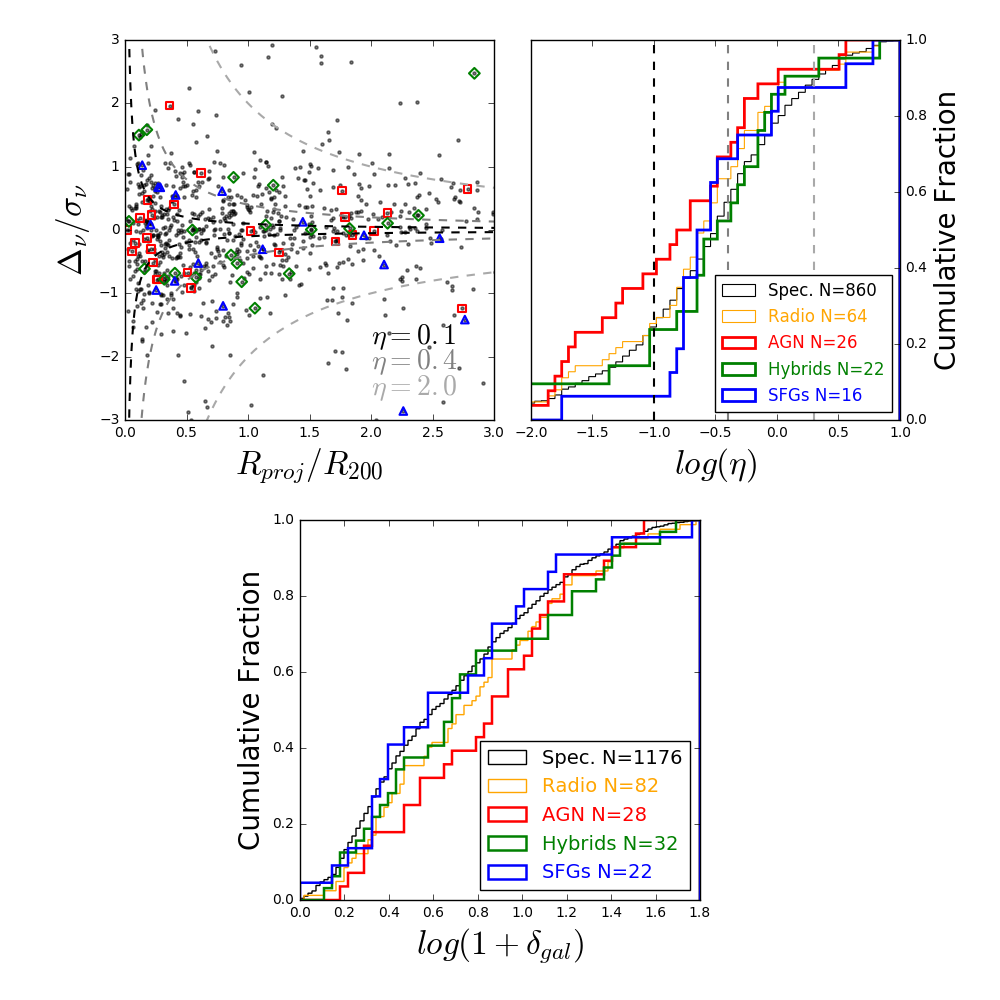}
    \caption{\textit{Top left}: $R_{proj}/R_{200}$ versus $|\Delta_\nu|/\sigma_\nu$ phase space diagram for spectroscopically-confirmed members (grey dots) and three radio populations (marked by open coloured squares). Three lines of constant $\eta$ are displayed as well. \textit{Top right}: CDFs of $\eta$. The red, green and blue histogram correspond to AGN, Hybrid and SFG population. The brown thin line represent the distribution of the overall radio sample, and the black line represents the distribution of spectroscopically-confirmed members. The number of galaxies in each class in both plots is displayed after their label. Three lines of constant $\eta$ are displayed with same colour in the top panel. \textit{Bottom}: CDFs of overdensity, on radio sub-classes, the overall radio and spectroscopically-confirmed samples. The colour convention is the same as the top right panel.}
    \label{fig:pp}
\end{figure*}

We derived $\eta$ for each galaxy relative to the center of every cluster/group in each LSS and select the smallest $\eta$ to determine the host cluster/group. We plot each galaxy population in the phase space diagram and $\eta$ distribution shown in the top two panels of Figure~\ref{fig:pp}, along with three lines of constant $\eta$. In the phase space diagram, the three sub-classes are marked with coloured open squares depending on their radio galaxy types, along with all spectroscopically-confirmed members in the five LSSs. However, the trend of each sub-class is not clear in this figure. 

To obtain a more quantitative look at the distribution on each sub-class as well as the overall trend of radio and spectroscopic samples, we plot the $\eta$ cumulative distribution functions (CDFs) (see Figure~\ref{fig:pp} top right panel) for objects with $R_{proj}/R_{200} < 3$ and $|\Delta_\nu/\sigma_\nu| < 3$ in the three radio sub-classes, in the overall radio sample and in the full spectroscopically-confirmed sample. The reason for applying these cuts is to include galaxies that are clearly associated with an individual cluster/group, since the possibility of a galaxy interacting with a cluster/group beyond the two radial/velocity cuts is small.

We find that the distribution of AGN increases rapidly below $\eta < 0.1$. On the contrary, the distribution of SFG grows significantly at $0.1 < \eta < 0.4$. Meanwhile, Hybrid galaxies are distributed similarly to that of the overall radio sample, with a rapid increase in the infall region at $0.4 < \eta < 2.0$. Again, their distribution does appear to resemble a mixture of AGN and SFGs, strongly consistent with the conclusion in the colour, stellar mass, radio luminosity and spectral analyses. 

We use the K-S test on the distributions of $\eta$ between AGN and SFGs, and find a p-value of $\sim0.10$, which implies that the AGN and SFGs do not share the same global environment distribution. Therefore, we conclude that AGN preferentially reside in the cluster/group environment, while SFGs generally avoid these regions. The K-S test between AGN and Hybrids, SFGs and Hybrids, and radio confirmed galaxies and spectroscopically-confirmed members are not conclusive (see Table \ref{tab:ks}). 
We perform the same sampling algorithm here as in Section~\ref{sec:cc}. We randomly select a sample from the combined AGN and SFG samples for 100 trials with the same number of galaxies in the Hybrid sub-class. In this way, we confirm that mixing AGN and SFGs can likely not produce the Hybrid distribution, as the mode of p-value $\sim0.09$, when comparing these composite distribution with these of the Hybrids.

\paragraph{Local Environments}
\label{sec:loc}
In the bottom panel of Figure~\ref{fig:pp}, we plot the CDFs of $log(1+\delta_{gal})$ of AGN, Hybrids and SFGs, along with the overall radio sample and spectroscopically-confirmed members. We only include galaxies at $log(1+\delta_{gal}) \geq 0$ as the spectroscopic completeness of objects with $z_{phot}$ estimates consistent with the LSS redshift ranges and subject to the absolute/apparent magnitude and $M_{\ast}$ cuts imposed for all analyses at these overdensities is $\sim$40\% across all fields. At such levels of completeness we can be reasonably sure the spectroscopic LSS members trace the true member population. This completeness begins to drop precipitously if we include similar objects at lower overdensity values. 
We use the K-S test on the $log(1+\delta_{gal})$ distributions and obtain a p-value of $\sim0.10$ between the AGN and Hybrids and a p-value of $\sim0.24$ between the AGN and SFGs. The former result indicates that the two distributions are marginally different, suggesting that AGN preferentially reside in denser local environments than Hybrids. This result is, at least in part, a consequence of AGN preferentially inhabiting cluster/group cores. 
Hybrids and SFGs have a high confidence ($\sim 91\%$) of being drawn from the same distribution. We test whether the Hybrid class is a mixture of AGN and SFGs by performing the same sampling algorithm as in the Section~\ref{sec:cc}. The result of the p-value mode ($\sim 0.08$) supports the supposition that Hybrid galaxies are their own unique class as was the case when a similar test was performed on their global environments (see Section~\ref{sec:global}). The K-S test on the $log(1+\delta_{gal})$ distributions between the overall radio population and spectroscopic members is not conclusive.

Combining the results from the global and local environment analyses, AGN highly prefer the densest, cluster core regions consistent with other studies (e.g. ~\citealp{Best2000}). This result is expected since galaxies with older stellar populations and suppressed star-formation are found in virialized cluster core regions (e.g. ~\citealp{Noble2016}). 

SFGs prefer the outskirts of the cluster/group and poor local environments, as expected from studies of star-forming galaxies in the mid-infrared (e.g. ~\citealp{Starikova2012}), FIR (e.g. ~\citet{Hickox2012}, and the UV band (e.g.~\citealp{Heinis2007}). 
The difference in the clustering properties of AGN and SFGs selected in the radio is also found in~\citet{Magliocchetti2016}, who classified radio galaxies based on their radio-luminosity at $0 \le z \le 4$, using the projected two-point correlation function $\omega(\theta)$ and the real space correlation function $\xi(r) = (r/r_0)^{-\gamma}$ as the cluster property parameters. They concluded that AGN are more clustered than SFGs, consistent with our results. 

\subsubsection{Radio and Non-radio Spectroscopic CMC Sample Comparison}
\label{sec:control}
It is interesting to investigate whether the radio emission is triggered because of the host environment or its internal properties.  For example, studies have suggested that radio-loud AGN occupy richer environments than similarly massive radio-quiet galaxies at low (e.g.,~\citealp{Best2004, Magliocchetti2004, Kauffmann2008, Bardelli2010, Donoso2010}) and even high redshift~\citep{Hatch2014}. 
In addition, we find that, in Section \ref{sec:radio_spatial}, the K-S test on both global and local environment of the full radio sample compared to the spectroscopically confirmed members indicate that we cannot reject the hypothesis that the two distributions are drawn from the same distribution. To better probe any differences, we examine here the environmental dependence on the three radio sub-classes compared with spectroscopic members without radio emission that are matched in both colour and stellar mass. 
We obtain three colour-mass-control (CMC) non-radio spectroscopically-confirmed samples, following the methodology described in Appendix~\ref{app:control}, matched separately to the three radio sub-classes (named AGN-CMC, Hybrid-CMC and SFG-CMC). We sample the control sample in 100 trials to eliminate the bias from a small number of random selections. Again, we adopt $\eta$ and $log(1+\delta_{gal})$ as our metrics for global and local environment. 

\begin{figure*}
	 \includegraphics[width=0.8\textwidth]{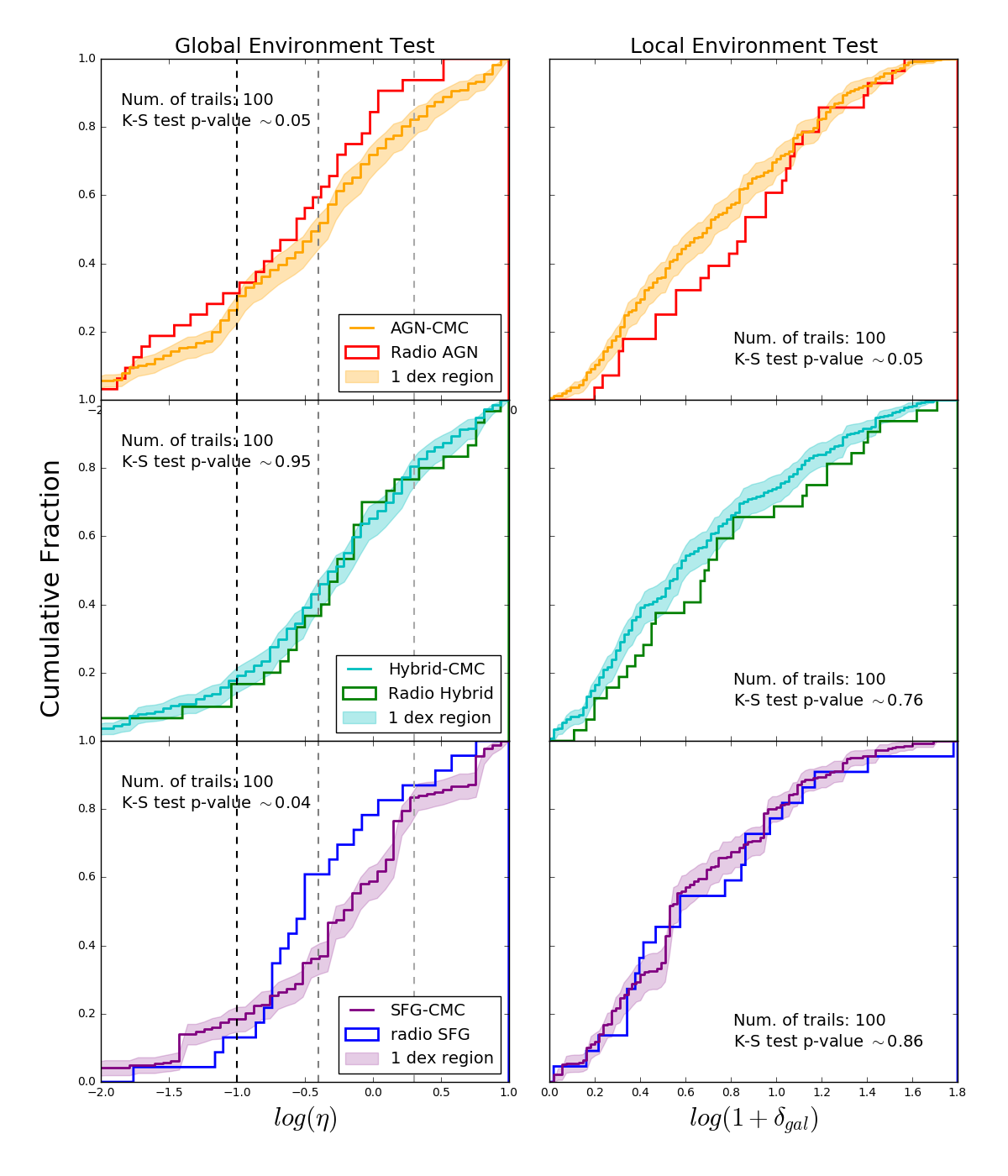}
	\caption{Global and local environment Cumulative Distribution histogram (CDF) on the radio sub-classes and their CMC sample.~\textit{Left}: CDF of $\eta = R_{proj}/R_{200} \times |\Delta_\nu|/\sigma_\nu$, for the three radio sub-classes and their CMC samples are plotted in the three panel, AGN (top), Hybrid (Middle), and SFG (Bottom), along with a $1\sigma$ shaded region on the mean of 100 re-samples.~\textit{Right}: CDF of $log(1 + \delta_{gal})$, for the three radio sub-classes (colour depends on their radio type) and their CMC samples are plotted in the three panel, along with a $1\sigma$ shaded region on the mean of 100 re-samples. Three lines of constant $\eta$=0.1, 0.4, 2.0 are displayed as well. }
	\label{fig:control}
\end{figure*}

We plot the $\eta$ and $log(1 + \delta_{gal})$ cumulative distributions of the three radio sub-classes and their 100 trails CMC samples in Figure~\ref{fig:control}. From the left three panels which show the $\eta$ distribution, we see that the AGN-CMC and SFG-CMC samples are significantly different from their parent radio populations, an impression which is confirmed with the K-S test. The AGN have a strong preference to be in the cluster/group centre and SFGs tend to dominate in the intermediate region, that is the outskirts of the cluster/group. The AGN and SFG results suggest that the radio emission is triggered by the global environment, rather than the galaxy colour and stellar mass properties. The Hybrid sub-class highly overlaps with the Hybrid-CMC sample, confirmed by the very high p-value from the K-S test, suggesting environment does not play a strong role for this sub-class. 

In the right panels which show the local overdensity distribution, the AGN-CMC sample appears significantly different from the AGN population, which is confirmed by the K-S test. The AGN have a strong preference to be in denser local environments, which is likely a consequence of their preferentially inhabiting cluster/group cores. Even though there are slight differences in the Hybrid and SFG panels, the K-S tests are not conclusive. Further analyses, such as splitting SFGs/Hybrids by radio luminosity or SFR, the former of which we attempt in section \ref{sec:agn} with radio AGN, may help constrain the effect of local environment on these radio sub-samples, although these tests would require larger sample size. 

The result is consistent with other studies of radio AGN. 
In an environmental analysis based on a 1 Mpc radius as a local environmental metric,~\citet{Malavasi2015} concluded that radio AGN are always located in local environments that are significantly richer and denser than those around galaxies in which radio emission is absent. Moreover,~\citet{Bardelli2010} found that only the radio AGN between $0 < z < 1$ have a significantly different environmental distribution based on control samples from the 4.5-8.0$\mu$m infrared colour-specific star-formation plane, while radio SFGs did not show any preference.~\citet{Hatch2014} found a significant tendency for radio-loud AGN to be in denser Mpc-scale environments by comparing radio-loud AGN with mass and redshift matched non-radio galaxies at $1.3 < z < 3.2$. We will discuss the possible reasons for the environmental dependence of AGN and SFGs in the section below.


\section{Discussion}
\label{sec:discussion}
We studied radio galaxies and their hosts in five large-scale structures at $0.65 \le z \le 0.96$ in the ORELSE survey. We obtain 89 radio member galaxies in the LSSs by using VLA 1.4GHz observations down to a $4\sigma$ detection flux density limit of about $30 \mu Jy$, matched to our deep DEIMOS spectroscopic data using a maximum likelihood technique.  
 
We began this investigation by comparing the colour, stellar mass, radio luminosity, spectral properties and environmental preference of the three radio sub-classes: AGN, Hybrid and SFG, classified based on the two-stage radio classification which combines radio luminosity and colour-SRL criteria (see Section~\ref{sec:classification}). 
We showed that AGN, on average, were situated in the quiescent colour-colour region and are most massive in their stellar content of all radio sub-classes. Radio AGN have higher radio luminosity and have older hosts than those of SFG and Hybrid galaxies. AGN tend to be located in dense environments in general, even when compared to the AGN-CMC sample. On the other hand, SFGs were situated in the dusty star-forming colour-colour region and exhibited spectral features indicating of ongoing star-formation. SFGs preferred intermediate regions of clusters/groups and moderate local environments. The preference for intermediate global environment holds when comparing to the SFG-CMC sample. Hybrids appear in the intermediate region between AGN and SFGs in the colour-SRL diagram, however, we see differences of this sub-class with respect to other sub-classes in all the galaxy property analyses. Hybrids extend farthest from the colour-colour quiescent, star-forming regions separation line and have the least stellar mass of all radio sub-classes, and exhibited the strongest [\ion{O}{ii}], [\ion{Ne}{iii}] and H$\gamma$ emission in the average host galaxies of all three radio sub-classes. Hybrid hosts share the same environmental preferences as the overall radio population and also the Hybrid-CMC sample. Almost all the properties suggest that Hybrids are not simply an intermediate population between the SFG and AGN population, the exceptions being their intermediate stellar age and moderate dust extinction. 

\subsection{AGN}
\label{sec:agn}
All analyses support that the AGN host galaxies are ``red-and dead" at the oldest evolutionary stage, when star-formation is truncated, a result in line with other studies (e.g.~\citealp{Croton2006, Best2006}).
~\citet{Magliocchetti2016} studied a large sample of field radio AGN at $z \sim 1$, classified based on radio luminosity. The average stellar mass for the AGN population is $\langle M_*\rangle = 10^{10.9 \pm 0.5}~M_\odot$, slightly lower than our result of $\langle M_*\rangle = 10^{11.16 \pm 0.03}~M_\odot$. This result may suggest that the AGN hosts in LSSs have slightly higher stellar masses to those in the field. We note that our result is not due to a higher stellar mass limit in our survey (see Section~\ref{sec:sm}).  

Comparing to a control sample of non-radio spectroscopically-confirmed members matched in 3D colour and stellar mass phase space, the radio AGN were preferentially found to reside in the cluster/group core and denser local environments, with the local environment preference in line with some previous studies~\citep{Bardelli2010, Hatch2014, Malavasi2015}. This result is, however, in contrast to \citet{Hickox2009} in which radio AGN were compared to a control sample of non-radio emitting galaxies broadly matched in colour, absolute magnitude and redshift, and no difference was found in clustering in terms of the local environment. However, we note that they only included radio AGN with $log(L_{1.4GHz}) \geq 23.8$, which would include only 64\% of our radio AGN sample. This may hint that the difference in the local environment is driven mostly by low power radio AGN, a result also found in~\citet{Malavasi2015}. We perform the same analysis as shown in Section~\ref{sec:control} but splitting radio AGN into high and low power, using the same radio luminosity threshold ($log(L_{1.4GHz}) = 24$) as in~\citet{Malavasi2015}. Nevertheless, neither of these samples shows significant differences compared to their CMC samples respectively in global or local environment. This result is due to the small sample size of each radio AGN sub-sample. As we explore the radio AGN in the full sample of the ORELSE survey, we will examine these differences with larger sample sizes. Regardless, if true, the preference for higher local densities potentially allows for an increased chance, relative to the CMC sample, at triggering AGN activity via galaxy-galaxy interactions and the accretion of gas from companion galaxies.

It is believed that in field environments radio AGN are fueled primarily by the cooling of hot gas in the interstellar and intergalactic medium (\citealp{Heckman2014}, and reference therein). However, in our sample we probe largely intermediate- and high-density global environments and radio AGN hosts which are more massive in their stellar content than their field counterparts. Therefore, the mechanism exciting or maintaining the radio AGN mode for the galaxies in our sample may be different.
~\citet{Best2006} found that, in nearby clusters/groups, Brightest Cluster/group Galaxies (BCGs) are more likely to host a radio-loud AGN than other galaxies of the same stellar mass, which suggests that radio AGN could be triggered by a strong cooling flow in the hot envelopes of galaxies in the central cluster galaxies and the intracluster medium (ICM). They argued that Bondi accretion~\citep{Bondi1952} could be the possible mechanism.
~\citet{Balmaverde2008}, who extended the analysis performed by~\citet{Allen2006}, estimated the Bondi accretion rate of hot gas by using X-ray observations to derive profiles of the X-ray brightness and gas temperature. They suggested that Bondi accretion is the dominant process in radio AGN in the local universe. 
Recently,~\citet{Fujita2016}, by studying a sample of local BCGs, also supported the Bondi accretion to be a route for the gas supply to the central SMBHs in the BCGs. 
We use Most Massive Cluster Galaxies (MMCGs) to represent the BCG population following the method described in~\citet{Ascaso2014}. In brief, spectroscopic members for each cluster/group are defined as $R_{proj} \le 1$ Mpc and $|\Delta_v| \le 3\sigma_\nu$. Galaxies with stellar masses within $1\sigma$ of the highest stellar mass were also selected (setting a maximum of three candidate MMCGs for each cluster/group) as candidate MMCGs. We obtain 6 AGN, 2 Hybrids and zero SFG out of total 25 candidate MMCGs for the entire sample presented here. AGN, therefore, have a higher probability to be candidate MMCGs than the other two sub-classes. 
Therefore, we suggest that AGN activity could be ignited by the interaction with the hot gas in the dense medium as radio AGN host galaxies settling into the cluster/group core. However, the sample size is too small to compare with other studies, though we do have X-ray observation for  some clusters we examined here. As we explore the radio AGN in the full sample of the ORELSE survey, we will examine this scenario. While such a picture appears consistent with the bulk of the radio AGN hosts, radio AGN that inhabit less dense environments may need a different scenario.

\subsection{SFG}
\label{sec:sfg}
Properties examined in this paper indicate that SFGs are dusty starburst galaxies, preferentially found in intermediate regions with $0.1 < \eta < 0.4$. We found that our result of $\langle SFR_{1.4GHz} \rangle = 60.5~M_\odot yr^{-1}$ is consistent with~\citet{Bonzini2015}, in a similar depth radio survey of a field which spanned a wide redshift range ($0.1 < z < 4.0$). 
However, our SFG hosts are generally more massive that their sample, which suggests that radio SFGs in LSS may be different from those in the field.

~\citet{Webb2013} argued that star-forming galaxies avoid the highest density regions of the universe in a cluster survey at $z~\sim~0.75$. This result is in line with~\citet{Noble2013} who found that the specific star formation rate (sSFR) of member galaxies increases from the virtualized region to the intermediate region. In addition,~\citet{Bell2005} found galaxies with higher sSFR are more likely to be undergoing galaxy-galaxy merging. 
\citet{Kocevski2011} found that the 24$\mu$m detected cluster/group galaxies in the LSS of SC1604, one of which we study here, have a burst of star formation in the recent past compared to those in the field. They suggested that the galaxy-galaxy merging could be a possible driving mechanism as these galaxies were preferentially found in the intermediate region. Our radio detected SFGs, also found primarily in the intermediate region, support that galaxy-galaxy mergers could be the triggering mechanism.

As star formation is truncated in radio SFGs, they will turn into post-starburst galaxies, known as K+A galaxies which preferentially reside in cluster environments at both low and high redshift~\citep{Dressler1983, Dressler1999, Poggianti1999, Poggianti2006}. 
In a study of two ORELSE LSSs, one of which we study here (SG0023),~\citet{Lemaux2016} characterized the true post-starburst population in these LSSs finding that it consisted of both traditionally-selected K+A galaxies and a set of K+A galaxies with [\ion{O}{ii}] emission originating from processes other than star formation, referred to a ``KAIROS" galaxies. 
When comparing the SFG population in this paper with the true K+A population from~\citet{Lemaux2016}, we find that both tend to be located in the intermediate region. Therefore, we propose a toy scenario in which SFGs will evolve into true K+As within the lifetime of a starburst ($\sim100-500$ Myr;~\citealp{Swinbank2006, Hopkins2008, McQuinn2009, McQuinn2010, Wild2010}). This time period is not long enough for these galaxies broadly change their global environments (max change $0.3 \sim 0.4$ Mpc).
We derive an evolved SFG stellar mass distribution that evolves in two phases. Firstly, we consider that SFGs will keep a constant SFR, derived from the radio luminosity\footnote{Adopting $SFR_{1.4GHz}$ equation from~\citet{Bell2003} and multiplying the derived SFR by a factor of 0.6 to convert from a Salpeter to a Chabrier IMF.}, for another $100~Myr$. 
Secondly, we assume 25\% to 80\% stellar mass loss, due to a major merging event (e.g.,~\citealp{Conroy2007, Murante2007, Rudick2011}) and stellar feedback (\citealp{Hopkins2014} and references therein).
We run a Monte-carlo simulation by drawing a random mass loss ratio between 25\% to 80\% for 100 trials. We utilize the K-S test to compare the evolved stellar mass distribution of each trial with that of the true K+A population. Large p-values in most of the cases confirm that this is a plausible scenario, as long as radio SFGs follow this scenario with a certain mass loss ratio (25\% to 80\%), though this restriction is obviously degenerate with the form of the assumed star formation history.  

Ram-pressure stripping (RPS)~\citep{Gunn1972, Quilis2000, Treu2003} could be a quenching mechanism that truncates our starbursting SFGs, so that they become the K+A population that we have observed in ORELSE. 
~\citet{Noble2016} found, in a cluster survey, that the SFR of galaxies at $z\sim 1.2$ decreases surprisingly in $0.20 \le \eta \le 0.64$ where galaxies are susceptible to the RPS, regions which contain the vast majority ($\sim75\%$) of the radio SFGs in this study. 
However, the effectiveness of ram pressure stripping is highly dependent on both properties of the medium where the galaxy is sitting and the galaxy velocities relative to this medium~\citep{moran2007}, both of which vary immensely for the LSSs studied here~\citep{Rumbaugh2012, Rumbaugh2016}. 
Additionally, other studies from ORELSE suggest ram-pressure stripping events are not the primary cause of the K+A population~\citep{Wu2014, Lemaux2016} and thus preclude it as the main channel for quenching the radio SFG population here. Such lines of thought will be followed further when investigating the environments of SFGs and K+As in the full ORELSE sample over a wider redshift range of LSSs.

\subsection{Hybrid}
\label{sec:hybrid}
Throughout the analyses of Hybrids in the Section~\ref{sec:properties}, we argue that the Hybrid population is not a mixture of AGN and SFGs, but is rather driven by a different mechanism. Hybrid hosts are broadly distributed in colour offset (see Figure~\ref{fig:cc}) and stellar mass (see Figure~\ref{fig:mass}), and tend to be found in cluster/group infall region. However, such environmental preferences do not persist when compared to their CMC sample. They have ongoing star-formation and younger stellar ages than the AGN, but older than SFGs. These properties are reminiscent of another well-known galaxy population. 

Radio AGN are often classified into high-excitation radio galaxy (HERG) and low-excitation radio galaxy (LERG), based on the existence of high-excitation (HE) emission lines in the optical spectra of their host galaxies (e.g.~\citealp{Hine1979, Laing1994, Condon2002, Mauch2007, Smolcic2008, Padovani2009}). HERGs show an intermediate stage between being star formation dominated and AGN dominated with high accretion efficiency (e.g. ~\citealp{Smolcic2009a, Smolcic2009b, Moric2010, Hardcastle2013, Best2012}). 

As we have shown in Section~\ref{sec:spectra}, stronger [\ion{O}{ii}], [\ion{Ne}{iii}] and H$\gamma$ emission was observed for the average Hybrid host galaxies as compared to the other two radio sub-classes. Here we quantify this allegation. 
We re-measure the EW of [\ion{O}{ii}] and H$\gamma$ using a double Gaussian fit, including the stellar continuum absorption for the latter feature. 
We also measure the EW of [\ion{Ne}{iii}] using this fitting method but with a single Gaussian model. 
From the measured strength of H$\gamma$ feature and under the assumption of Case B recombination and $T_e = 2 \times 10^4 K$ and $n_e = 100 cm^{-3}$, 
we derive an $EW([\ion{O}{ii}])/EW(H\alpha)$ of $2.39 \pm 0.15$.
This value falls in the low-$[\ion{O}{ii}]/H\alpha$ region (see Figure 2 of~\citealp{Yan2006}), which is occupied by star-forming galaxies and type-2 Seyferts as shown in Figure 11 of~\citet{Lemaux2010}. The latter population has a large overlap with HERGs (e.g.,~\citealp{Donoso2009}).
In addition, we measure $EW([\ion{Ne}{iii}])/EW([\ion{O}{ii}])$ of $0.066 \pm 0.005$.
This ratio, in conjunction with estimated restframe colours, can be used to determine the dominant emission mechanism~\citet{Trouille2011}. Here we use $M_V-M_J$ instead of $M_g-M_z$. The Hybrid population is found to be comfortably situated in the AGN region.
In a similar approach, when combining $EW([\ion{Ne}{iii}])/EW([\ion{O}{ii}])$ ratio with the average D(4000) of Hybrids, we find that the population is consistent with Seyferts/composites and completely inconsistent with having emission driven by pure star formation 
(see Figure 3 in ~\citealp{Marocco2011}). 

All of the evidence supports the hypothesis that Hybrids are a HERG population. 
Meanwhile, Hybrids tend to be located in similar global and local environments as SFGs, which indicate that Hybrids could be affected by galaxy-galaxy interactions or merging activities~\citep{Kocevski2011, Kartaltepe2012, Pawlik2016}. 
It is clear that Hybrids are not the same population as either radio AGN or SFGs.  
We will continue to study this population as we investigate the full ORELSE survey in a wider redshift range.

\section{Conclusions}
\label{sec:conclusion}
In this work, we study the three radio sub-classes, using a two stage classification scheme based on the radio luminosity and the colour-SRL. In particular, we investigate the colour, stellar mass, radio luminosity, average host spectra and the effect of both global and local environments for each radio sub-class. 
For the first time, we compare each radio sub-class with a three-dimensional ($M_{NUV}-M_{r}$, $M_{r} - M_{J}$ and stellar mass) phase space controlled non-radio spectroscopically-confirmed sample on both global and local environments. 
Our main conclusions are the following:
\begin{itemize}
\item AGN hosts are the most massive population and exhibit a quiescence in their star-formation activity. They tend to be located in the cluster/group virialized region and locally dense environments with both preferences persisting when compared to its CMC sample. These preferences suggest that the AGN host galaxies in the cluster/group cores have their AGN ignited by interactions with the dense medium and possibly by increased interactions with neighboring galaxies.

\item The SFG population is in the star-forming colour-colour region and have comparable stellar masses as those galaxies hosting a radio AGN. The coadded optical spectrum reveal them to be dominated by young stars. SFGs tend to be in the intermediate region in both the global and the local environments with the former preference persisting when contrasted to its CMC sample. This and other evidence suggest a connection to a galaxy merging origin. 
As SFGs truncate the star-formation process, they will turn into post-starburst galaxies. This scenario is supported by comparing the stellar mass distribution of radio SFGs and the most recently post starburst galaxies in ORELSE.

\item Hybrids, though selected as an intermediate region in the colour-SRL, were found in almost all analyses to be a unique type of radio galaxies rather than a mixture of AGN and SFGs. The spectral analyses strongly suggest they have coeval star-formation and AGN activity with high accretion efficiency.

\item In addition, comparing to previous published studies of field radio galaxies, we find radio galaxies in LSSs are more massive, which suggest a mass-driven phenomena in LSSs. As we compare the field versus the LSSs in ORELSE, we will investigate this phenomena in an attempt to determine whether it is a global or a local effect and how it affects host galaxies.

For future studies we will include radio galaxies from all 16 ORELSE fields with VLA/JVLA obvservations which will allow us to triple our LSS sample and to construct an internal field sample of comparable size. In addition to the increased numbers, these samples will span a considerably larger range in redshift ($0.65 < z < 1.3$), allowing us separate out redshift-dependent effects from those originating from secular or environmental processes. With these larger samples, we should be able to definitively confirm both global and local environmental preferences of various radio sub-classes across a wider dynamic range of environments and understand whether such preferences evolve as a function of cosmic epoch.

\end{itemize}

\section*{Acknowledgements}

This material is based upon work supported by the National Science Foundation under Grant No. 1411943. Part of the work presented herein is supported by NASA Grant Number NNX15AK92G. 
This study is based on data taken with the Karl G. Jansky Very Large Array which is operated by the National Radio Astronomy Observatory. The National Radio Astronomy Observatory is a facility of the National Science Foundation operated under cooperative agreement by Associated Universities, Inc. 
This work is based, in part, on data collected at the Subaru Telescope and obtained from the SMOKA, which is operated by the Astronomy Data centre, National Astronomical Observatory of Japan; observations made with the Spitzer Space Telescope, which is operated by the Jet Propulsion Laboratory, California Institute of Technology under a contract with NASA; and data collected at UKIRT which is supported by NASA and operated under an agreement among the University of Hawaii, the University of Arizona, and Lockheed Martin Advanced Technology centre; operations are enabled through the cooperation of the East Asian Observatory. When the data reported here were acquired, UKIRT was operated by the Joint Astronomy Centre on behalf of the Science and Technology Facilities Council of the U.K. 
This study is also based, in part, on observations obtained with WIRCam, a joint project of CFHT, Taiwan, Korea, Canada, France, and the Canada-France- Hawaii Telescope which is operated by the National Research Council (NRC) of Canada, the Institut National des Sciences de l'Univers of the Centre National de la Recherche Scientifique of France, and the University of Hawai'i. The scientific results reported in this article are based in part on observations made by the Chandra X-ray Observatory and data obtained from the Chandra Data Archive. 
The spectrographic data presented herein were obtained at the W.M. Keck Observatory, which is operated as a scientific partnership among the California Institute of Technology, the University of California, and the National Aeronautics and Space Administration. The Observatory was made possible by the generous financial support of the W.M. Keck Foundation. 
We wish to thank the indigenous Hawaiian community for allowing us to be guests on their sacred mountain, a privilege, without with, this work would not have been possible. We are most fortunate to be able to conduct observations from this site. 




\bibliographystyle{mnras}
\bibliography{reference}  



\appendix
\section{Control Sample Selection}
\label{app:control}

Here we describe the process of selecting non-radio spectroscopically-confirmed colour and stellar mass control sample used in Section~\ref{sec:control}. The goal of these control samples is to eliminate colour and stellar mass differences between radio and non-radio normal galaxies, so as to investigate the environmental effects on igniting radio emission, independent of colour and stellar mass effects. 
We use the AGN population as an example parent-sample to explain this method. The same methodology is applied to the SFG and Hybrid sub-classes.
\begin{itemize}
\item Remove radio galaxies out from spectroscopic sample to obtain a non-radio population; 
\item We choose $M_{NUV}-M_{r}$, $ M_{r} - M_{J}$ and stellar mass to be the control parameters, and set them to be x-, y-, z-axis, respectively, in a 3D phase space. We then split the 3D phase space into $4 \times 4 \times 4$ boxes, which cover the full parameter space. If the box and therefore the control sample size is too small, we may be biased by large variations. 
\item We obtain a 3D probability density map by taking a ratio of the number of galaxies from the parent sample in each box over the total number of galaxies in the parent sample (N). However, if there is only one spectroscopic object in a box, we set the box density to be zero to avoid drawing one spectroscopic member multiple times in each trail.
\item We randomly draw from a sample of non-radio galaxies without replacement based on the 3D probability density map described above, with the same size of the parent sample, and designate the resultant sample as the Colour Mass Controlled (CMC) sample. We apply the K-S test to the CMC sample and the parent sample for the $\eta$ and $log(1 + \delta_{gal})$ distributions. 
\item The above process is repeated 100 time to obtain 100 CMC samples with 100 p-values for each of the comparisons to $\eta$ and $log(1 + \delta_{gal})$. We calculate the mode of the p-values by binning them into 10 bins and returning the bin with the most p-values. The mode of p-values is then the median of p-values in this bin. 
\end{itemize}

\section{Representativeness of the spectroscopic LSS Members sample}
\label{app:spec-photo}

Because the five LSSs have variable spectroscopic completeness brighter than $i^{\prime} \le 24$, it is possible that some radio hosts have been missed by our spectroscopic coverage. Further, since our initial spectral masks tended to target redder galaxies, we here examine the spectroscopic representation of the underlining overall galaxy population in the LSSs. We choose colour $M_{NUV}-M_{r}$ and stellar mass $log(M_*/M_\odot)$ to test the spectroscopic representation, to determine any bias which may affect our analysis in the Section~\ref{sec:properties}, especially in Section~\ref{sec:control}, where we compare the environmental preference of the radio confirmed sample to all spectroscopic members. 

We apply two selection methods to recover the underlining overall LSS galaxy populations:  $z_{phot}$ selection and P(z) selection. To be clear, these selections are only for photometric objects without spectroscopic redshifts in the overall field ($non~z_{spec}~Phot$ sample). Then, we adopt two algorithms to test the spectroscopic representation: the spectroscopic ratio and the 2D K-S test to confirm the limits to which our spectroscopy is representative. 

The first selection method use $z_{phot}$, derived from SED fitting (see details in Section~\ref{sec:photoobs}), to select $non~z_{spec}~Phot$ sample in the $z_{phot}$ redshift range, where the $z_{phot}$ redshift range is given by 
\begin{equation}
	\begin{aligned}
    	 	z_{phot, min} &= z_{spec,min} - \Gamma \sigma \times (1 + z_{spec, min}) \\
	 	z_{phot, max} &= z_{spec,max} - \Gamma \sigma \times (1 + z_{spec, max})
	\end{aligned}
	\label{eq:photoz}
\end{equation}
with $\sigma$ is the standard deviation of a fitted Gaussian to the distribution of $(z_{spec} - z_{phot} )/(1 + z_{spec} )$ measurements in the range $0.5 < z <1.2$. $\Gamma$ is chosen to maximize the precision and recall of selecting spectroscopic objects on their estimated $z_{phot}$ compared to selection on $z_{spec}$. We find $\Gamma = 1$ to be the optimal value for all five fields. We call this method $z_{phot}$ selection and combine these objects with the spectroscopically-confirmed members to create a $z_{phot}-Phot$ sample. The number of $z_{phot}-Phot$ sample for each field is summarized in Table~\ref{tab:photo}. 

We also use the probability P(z) for each source described in Section~\ref{sec:photoobs} following the same method used in~\citet{Rumbaugh2017}, to estimate the probability of a source without a spectroscopic redshift to be a member of a given LSS by integrating P(z) over the spectroscopic redshift bounds of that LSS. We find the probability threshold used to select photometric members varies from field to field, mainly because the spectroscopic redshift window of the LSSs are significantly different among the five fields. Therefore, we calculate the precision and recall on the $non~z_{spec}~Phot$, depending on the estimated true population using P(z) probabilities (see Equation~\ref{eq:cp}) to find a reliable probability threshold to select $non~z_{spec}~Phot$ in the LSSs.  
\begin{equation}
	\begin{aligned}
    	 	Recall = \frac{\sum Prob_{non~z_{spec}~Phot} (> P_i) + N_{spec}} {\sum Prob_{non~z_{spec}~Phot} + N_{spec}}\\
   	 	Precision = \frac{\sum Prob_{non~z_{spec}~Phot.}  (> P_i) + N_{spec}} {N_{non~z_{spec}~Phot} (> P_i) + N_{spec}}
	\end{aligned}
	\label{eq:cp}
\end{equation}
where the $N_{spec}$ is the number of the spectroscopically-confirmed members sample, $\sum Prob_{non~z_{spec}~Phot} (> P_i)$ is the sum of the probability of $non~z_{spec}~Phot$ above $P_i$, which indicates the number of $non-z_{spec} Phot$ in the LSS, and $N_{non-z_{spec}~Phot.} (> P_i)$ is the total number of $non~z_{spec}~Phot$ given the chosen $P_i$. We call this method P(z) selection and combine the objects that meet this criterion with the spectroscopically-confirmed members to create a $P(z)-Phot$ sample. The number of $P(z)-Phot$ sample for each field is summarized in Table~\ref{tab:photo}. 

\begin{table*}
	\caption{Photometric members}
	\label{tab:photo}
	\begin{threeparttable}
	    \begin{tabular}{ l c | c c c || c c} 
		    \hline
		    \hline
		   & & \multicolumn{3}{ | c | }{$z_{phot}$ selection} & \multicolumn{2}{| c |}{P(z) selection} \\
		   \hline
	    	   LSS & Photometric  & $z_{phot}$  & $\sigma$ & $z_{phot}$ Phot. &  Probability & $P(z)$ Phot. \\
	    	    & Sources\tnote{1} & LSS Range\tnote{2} & & Members\tnote{3} & Threshold & Members\tnote{4} \\
		    \hline
		    SC1604 & 6622 & 0.78$\sim$1.02 & 0.029 & 1537 & 0.43 &957\\
	    	    SG0023 & 7431 & 0.77$\sim$0.93 & 0.025 & 1396 & 0.19 &448\\
		    SC1324 & 7511 & 0.60$\sim$0.84 & 0.033 &2156& 0.41 &1228\\
		    RXJ1757 & 2972 &  0.63$\sim$0.76 & 0.027 & 378& 0.18 &101\\
		    RXJ1821 & 1702 & 0.75$\sim$0.90 & 0.029 & 382& 0.10 &134\\ 
		    \hline
	    \end{tabular}
	    \begin{tablenotes}
	    	\item[1] Includes photometric detections with use flag, $18.5 \le i' \le 24.5$ and restricted to be in the same spectroscopic spatial area;
		\item[2] Photometric redshift boundaries defined for each LSS, using $\sigma$ in the fourth column, with $\Gamma = 1$;
		\item[3] Including photometric detections in the $z_{phot}$ LSS range, for $non~z_{spec}~Phot$, together with spectroscopically-confirmed members. 
		\item[4] Including photometric detections using $P(z)$ selection for $non~z_{spec}~Phot$, together with spectroscopically-confirmed members.
	    \end{tablenotes}
    \end{threeparttable}
\end{table*}

\begin{figure*}
    \includegraphics[width=\textwidth]{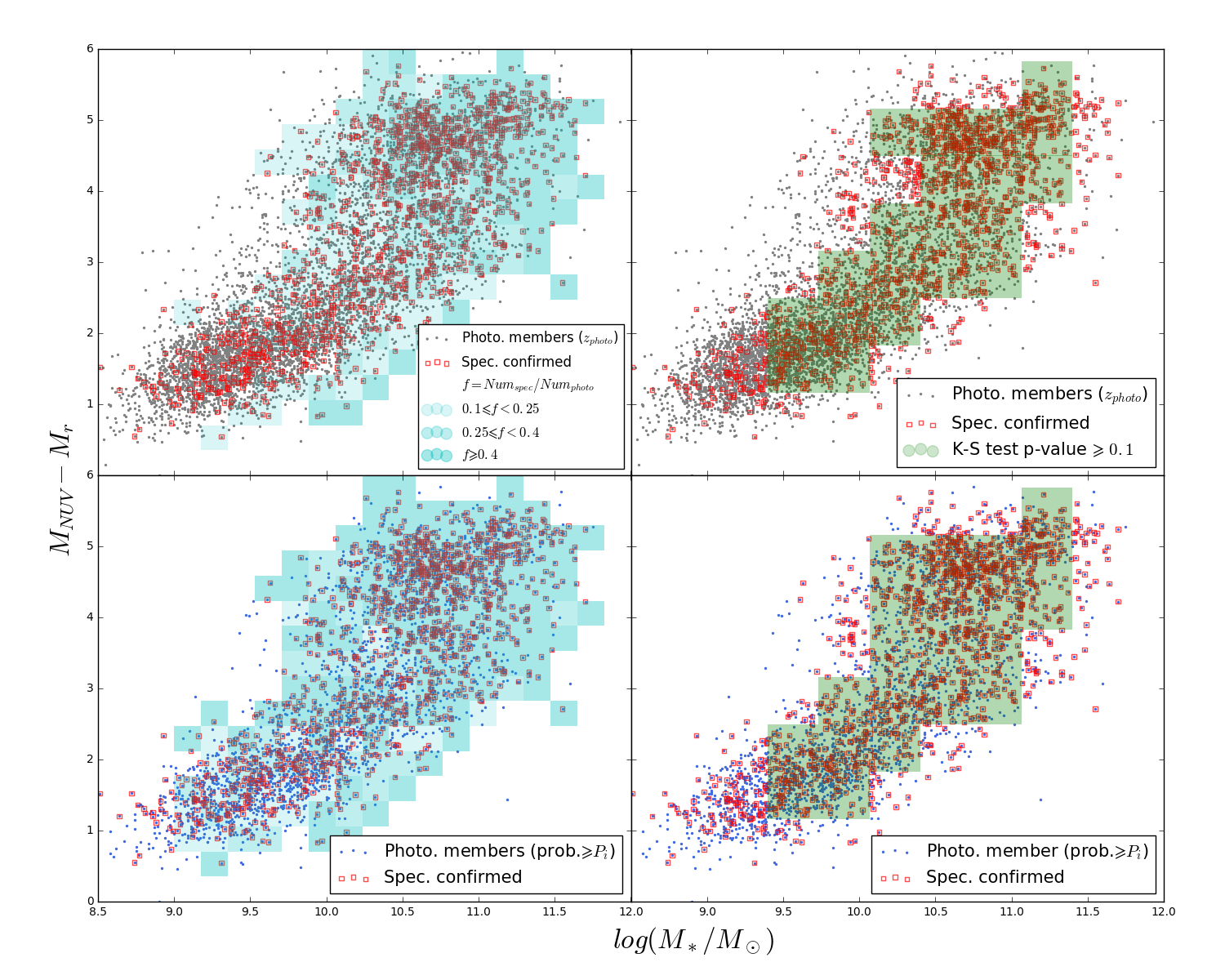}
    \caption{$M_{NUV}-M_{r}$ versus stellar mass diagram (colour-mass diagram). The red open squares are the spectroscopically-confirmed members. \textit{Top left}: Grey dots are photometric objects using $z_{phot}$ selection, with shaded regions of $f > 0.1$, the darker the colour, the higher the f;  \textit{Bottom left}: Blue dots are photometric objects using $P(z)$ selection, with same criteria for the shaded regions as the top left panel; \textit{Top right}: Grey dots are photometric objects using $z_{phot}$ selection, testing with 2D K-S test, with shaded regions of p-value $\geq 0.1$; \textit{Bottom right}: Blue dots are photometric objects using $P(z)$ selection, testing with 2D K-S test, with shaded regions of p-value $\geq 0.1$.}
    \label{fig:completeness}
\end{figure*}

To more robustly test the representativeness of our spectroscopy, we apply two algorithms to confirm the colour and stellar mass limit for the spectroscopic members. Firstly, we calculate a ratio of the number of spectroscopic members over the number of photometric members in either two selection methods, shown in the left two panels of Figure~\ref{fig:completeness} (top panel uses the $z_{phot}$ selection and bottom uses the $P(z)$ selection method), with the cyan shaded region representing the spectroscopic ratio. We find a constant ratio in each area for $M_{NUV}-M_{r} > 2.5$ and $M_* > 10^{10.2}~M_\odot$, though the ratio is generally lower in the top $z_{phot}$ selection panel than the bottom $P(z)$ selection panel. 

As the second algorithm, we adopt a generalization of a two-dimensional distributions from~\citet{Peacock1983} and~\citet{Fasano1987}. Its detailed implementation is described by~\citet{Press2007}. The 2D K-S test computes a p-value similar to the traditional 1D K-S test, which can be interpreted as the probability of two data sets being drawn from the same distribution, with smaller p-value indicating a smaller chance of them being drawn from the same distribution. We accept the two distributions with p-value above 0.1 to be likely from the same one and reject the non hypothesis if the number of objects is below 20, since, with such a small data set, the p-value from 2D K-S test is not reliable. We show the result in the right two panels of Figure~\ref{fig:completeness} again with the top panel using the $z_{phot}$ selection method and bottom panel using the $P(z)$ selection method. The green shaded region indicates where p-value $\le 0.1$. The 2D K-S test confirms that, in the region of $M_{NUV}-M_{r} > 2.5$ and $M_* > 10^{10.2}~M_\odot$, the two populations are statistically distinguishable. 

We combine the four results together as our conclusion. Therefore, the spectroscopic sample in the five LSSs are representative above $M_{NUV}-M_{r} > 2.5$ and $M_* > 10^{10.2}~M_\odot$, no matter which selection and test is used. In addition, we employ various tests (K-S test, t-test, Mann-Whitney U test) on the colour and stellar mass, separately, to spectroscopically-confirmed members and $z_{phot}-Phot$ sample (or $P(z)-Phot$ sample) which restrict to $M_{NUV}-M_{r} > 2.5$ and $M_* > 10^{10.2}~M_\odot$. The results confirm our conclusion that the spectroscopically-confirmed members are representative of the underlying true LSS member sample with $M_{NUV}-M_{r} > 2.5$ and $M_* > 10^{10.2}~M_\odot$. 

We note that the radio member sample would not be affected by the spectroscopic member sample not being representative of the true galaxy population, as discussed in Section~\ref{sec:cc} Figure~\ref{fig:cc} and Section~\ref{sec:sm} Figure~\ref{fig:mass}. Moreover, we show in the Section~\ref{sec:control}, that the CMC samples should also not be affected, because the control algorithm, described in Appendix~\ref{app:control}, only chooses non-radio spectroscopic samples weighted by the density of the radio sub-class which are not biased by an unrepresentation of the spectroscopic member sample. 

\section{The Available Imaging and Their Limits}
\label{app:images}
We list, in tables \ref{tab:images1} and \ref{tab:images2}, the available imaging for each field and the $5\sigma$ point source completeness limits for images in each band.
\begin{table}
	\centering
	\caption{Imaging Data}
	\label{tab:images1}
	\begin{threeparttable}
	\begin{tabular}{lll}
		\hline 
		Band & Telescope/Instrument & Depth\tnote{1}\\ 
		SC1604 & & \\
		\hline
		\hline
		$B$ & Subaru/Suprime-Cam & 26.4 \\
		$V$ & Subaru/Suprime-Cam & 25.9\\
 		$Rc$ & Subaru/Suprime-Cam & 26.0\\
		$r'$ & Palomar/LFC & 25.2\\
		$Ic$ & Subaru/Suprime-Cam & 25.0\\
		$i'$ & Palomar/LFC & 24.5\\
		$Z^+$ & Subaru/Suprime-Cam & 24.5\\
		$z'$ & Palomar/LFC & 23.2\\
		$J$ & UKIRT/WFCAM & 22.3\\
		$K$ & UKIRT/WFCAM &  22.0\\
		$[3.6]$ & \textit{Spitzer}/IRAC &  24.2\\
		$[4.5]$ & \textit{Spitzer}/IRAC & 23.9\\
		$[5.8]$ & \textit{Spitzer}/IRAC & 22.3\\
 		$[8.0]$ & \textit{Spitzer}/IRAC & 22.4\\
		\hline
		SG0023 & & \\
		\hline
		\hline
		$B$ & Subaru/Suprime-Cam & 26.4 \\
		$V$ & Subaru/Suprime-Cam & 25.9 \\
		$R^+$ & Subaru/Suprime-Cam & 25.2 \\
		$r'$ & Palomar/LFC & 25.1 \\
		$I^+$ & Subaru/Suprime-Cam & 24.6 \\
		$i'$ & Palomar/LFC & 24.5 \\
		$z'$ & Palomar/LFC & 23.1 \\
		$J$ & UKIRT/WFCAM & 22.0 \\
		$K$ & UKIRT/WFCAM & 22.0 \\
		$[3.6]$ & \textit{Spitzer}/IRAC & 22.2 \\
		$[4.5]$ & \textit{Spitzer}/IRAC & 21.9 \\ 
		\hline
		SC1324 & & \\
		\hline
		\hline
		$B$ & Subaru/Suprime-Cam & 27.2\\
		$V$ & Subaru/Suprime-Cam & 26.5\\
		$Rc$ & Subaru/Suprime-Cam & 25.8\\
		$r'$ & Palomar/LFC &  24.6\\
		$I^+$ & Subaru/Suprime-Cam & 24.6\\
		$i'$ & Palomar/LFC & 24.1\\
		$Z^+$ & Subaru/Suprime-Cam & 23.8\\
		$z'$ & Palomar/LFC & 24.1\\
		$J$ & UKIRT/WFCAM & 23.4\\
		$K$ & UKIRT/WFCAM & 22.7\\
		$[3.6]$ & \textit{Spitzer}/IRAC & 24.2\\
		$[4.5]$ & \textit{Spitzer}/IRAC & 24.2\\ 
		\hline
	\end{tabular}
    \begin{tablenotes}
    	\item[1] 5$\sigma$ point source completeness limit.
    \end{tablenotes}
\end{threeparttable}
\end{table}

\begin{table}
	\centering
	\caption{Imaging Data (continue)}
	\label{tab:images2}
	\begin{threeparttable}
	\begin{tabular}{lll}
		\hline
		Band & Telescope/Instrument & Depth\tnote{1}\\ 
		RXJ1757 & & \\
		\hline
		\hline
		$B$ & Subaru/Suprime-Cam & 26.8\\
		$V$ & Subaru/Suprime-Cam & 26.2\\
 		$Rc$\tnote{6} & Subaru/Suprime-Cam & 26.8\\
		$r'$ & Palomar/LFC &  25.8\\
		$i'$ & Palomar/LFC & 25.2\\
		$Z^+$\tnote{6} & Subaru/Suprime-Cam & 25.8\\
		$z'$ & Palomar/LFC & 24.0\\
		$Y$ & Subaru/Suprime-Cam & 23.5\\
		$J$ & CFHT/WIRCam & 21.6\\
 		$Ks$ & CFHT/WIRCam & 22.3\\
		$[3.6]$ & \textit{Spitzer}/IRAC & 22.6\\
		$[4.5]$ & \textit{Spitzer}/IRAC & 22.6\\ 
		\hline
		RXJ1821 & & \\
		\hline
		\hline
		$B$ & Subaru/Suprime-Cam & 26.5\\
		$V$ & Subaru/Suprime-Cam & 26.4\\
		$r'$ & Palomar/LFC &  25.7\\
		$i'$ & Palomar/LFC & 25.2\\
		$z'$ & Palomar/LFC & 24.1\\
		$Y$ & Subaru/Suprime-Cam & 24.0\\
		$J$ & CFHT/WIRCam & 22.1\\
		$Ks$ & CFHT/WIRCam & 22.4\\
		$[3.6]$ & \textit{Spitzer}/IRAC & 22.6\\
		$[4.5]$ & \textit{Spitzer}/IRAC & 22.6\\ 
		\hline
	\end{tabular}
\end{threeparttable}
\end{table}


\bsp	
\label{lastpage}
\end{document}